\documentclass{article}

\usepackage{arxiv}

\usepackage[utf8]{inputenc} 
\usepackage[T1]{fontenc}    
\usepackage{hyperref}       
\usepackage{url}            
\usepackage{booktabs}       
\usepackage{amsfonts}       
\usepackage{nicefrac}       
\usepackage{microtype}      
\usepackage{lipsum}
\usepackage{graphicx}
\graphicspath{ {./images/} }
\usepackage{epstopdf, epsfig}
\usepackage{caption}
\usepackage{color, soul}
\usepackage[colorinlistoftodos]{todonotes}

\usepackage{longtable} 
\usepackage{multirow} 
\usepackage{amsmath} 
\usepackage[T1]{fontenc} 
\usepackage{tabularx} 
\usepackage{changepage} 
\usepackage{natbib} 
\usepackage{float} 
\usepackage[running]{lineno} 
\usepackage{xcolor} 
\usepackage{relsize} 
\usepackage{url} 
\usepackage{appendix} 
\usepackage [english]{babel}
\usepackage [autostyle, english = american]{csquotes}
\MakeOuterQuote{"}
\title{\rm{Revisiting "bursts" in wall-bounded turbulent flows}}

\author{
Subharthi Chowdhuri \\
Department of Civil and Environmental Engineering\\
University of California, Irvine, CA 92697, USA \\
  \texttt{subharc@uci.edu} \\
   \And
Tirtha Banerjee \\
Department of Civil and Environmental Engineering\\
University of California, Irvine, CA 92697, USA \\ 
 \texttt{tirthab@uci.edu} \\
}

\begin{document}
\maketitle
\begin{abstract}
Turbulent signals are known to exhibit burst-like activities, which affect the turbulence statistics at both large and small scales of the flow. In our study, we pursue this problem from the perspective of an event-based framework, where bursting events are studied across multiple scales in terms of both their size and duration. To illustrate our method and assess any dependence on the Reynolds number ($Re$), we use two datasets - from the Melbourne wind tunnel ($Re \approx 14750$) and from SLTEST - an atmospheric surface layer experiment ($Re \approx 10^6$). We show that an index, namely the “burstiness index”, can be used successfully to describe the multi-scale nature of turbulent bursting while accounting for the small-scale intermittency effects. Through this index, we demonstrate that irrespective of $Re$, the presence of large amplitude fluctuations in the instantaneous velocity variance and momentum flux signals are governed by the coherent structures in the flow. Concerning small-scale turbulence, a $Re$-dependence is noted while studying the scale-wise evolution of the burstiness features of second-order streamwise velocity increments (${(\Delta u)}^2$). Specific to the wind-tunnel dataset, the burstiness index of ${(\Delta u)}^2$ signal displays a strong dependence on height and also decreases as the scales increase with the maximum being obtained at scales comparable to the dissipative structures. However, such features are nearly absent in the atmospheric flows. To conclude, this research paves a novel way to evaluate the effect of bursts on the turbulence statistics at any specified scale of the flow.
\end{abstract}


\section{Introduction}
\label{Intro}
In any stochastic signal, bursts are typically characterized by the presence of strong amplitude fluctuations, exceeding the standard deviation of the signal by multiple orders \citep{farazmand2017variational}. Understanding the origin of these bursts are important, since these are often known to occur in a plethora of physical systems. Some of their examples include (but not limited to): (1) extreme dissipation and flux events in turbulent flows \citep{yeung2015extreme,deshpande2021characterising}; (2) rogue waves appearing on sea surfaces \citep{dysthe2008oceanic}; (3) large solar flare events in astrophysical systems \citep{boffetta1999power}; (4) extreme rainfall events in weather and climate systems \citep{roxy2017threefold}, and so on. 

In the context of turbulence research, perhaps the first documentation of bursts was carried out by \citet{kline1967structure} while observing the occasional break-up of the near-wall streaks in wall-bounded turbulent flows. Typically, such bursting activities lead to large amplitude fluctuations in velocity variance and momentum flux signals, and therefore, they are considered to be an integral part of turbulence dynamics \citep{sapsis2021statistics}. Given their importance, since the study of \citet{kline1967structure}, numerous experimental and theoretical studies have been undertaken to understand the role of these bursts in turbulence production \citep{graham2021exact}. 

It is generally recognized that the presence of coherent structures, such as hairpin vortices, is primarily responsible for such bursting phenomenon \citep{robinson1991coherent,jimenez2018coherent}. Moreover, the researchers have shown that nearly 80\% of the Reynolds stress production happens through these bursts \citep{panton2001overview,farano2017optimal}. On the one hand, on the theoretical side, \citet{jimenez2013linear} has shown how the origin of these bursts can be explained through the solutions of Orr–Sommerfeld equations. On the other hand, in experimental research, the detection of bursts has mostly been achieved through variable interval time averaging (VITA) and quadrant-hole methods \citep{antonia1981conditional,morrison1988conditional,wallace2016quadrant}. Through such experimental schemes, one typically studies the dynamical features of extreme events in the instantaneous velocity variances and streamwise momentum flux signals, thereby connecting them with the coherent structures in the flow. In recent times, using direct numerical simulations, a few studies have been conducted to explore the three-dimensional topology of the coherent structures associated with these extreme events \citep{lozano2012three,dong2017coherent}.

From the above discussion, it is apparent that the bursts described so far are connected to the energy-containing structures (comparable to the integral scales) in the flow, and therefore, could be aptly characterized as large-scale bursts. However, in fully developed turbulent flows, there exist another type of bursts associated with smaller scales of the flow, comparable to the inertial subrange and dissipative range scales \citep{sreenivasan1997phenomenology}. These small-scale bursts are typically identified through the extreme events in velocity increments, such as in, $\Delta u(\tau)=u^{\prime}(t+\tau)-u^{\prime}(t)$, where $u^{\prime}$ is the streamwise velocity fluctuations, $t$ is time, and $\tau$ is the time-lag. 

In particular, the probability density functions of velocity increments become increasingly non-Gaussian as the eddy time scales decrease, a phenomenon associated with small-scale intermittency \citep{sreenivasan1997phenomenology}. The presence of such extreme events in the velocity increments disrupts the self-similarity of the small-scale eddy structures as predicted by Kolmogorov \citep{sreenivasan1997phenomenology}. This causes anomalous scalings in the higher-order structure functions ($\overline{{\vert \Delta u(\tau) \vert}^{m}} \neq m/3$, where $m$ is the moment order), which are often studied through multifractal analysis under the framework of phenomenological models such as the one proposed by \citet{she1994universal}. Moreover, the burst events in velocity increments ensure that the dissipation of turbulence kinetic energy remains finite even as the viscosity tends to zero, a feature commonly referred to as the dissipative anomaly \citep{frisch1995turbulence}.

As opposed to the non-Gaussianity associated with small-scale bursts, the fluctuating velocity signals through which the large-scale bursts are detected typically display near-Gaussian behavior \citep{morrison1988conditional}. Therefore, despite being fundamentally important, there lacks a unifying framework through which one can connect the small- and large-scale bursts. As a result, it remains largely unexplored how the burstiness features of a turbulent signal evolve as the scales of the eddies increase or decrease systematically. This issue is even more pertinent for high Reynolds number ($Re$) flows, which are characterized by a wide spectrum of eddy sizes. 

The recent reviews by \citet{graham2021exact} and \citet{sapsis2021statistics} show that the state-of-the-art theoretical models, mostly borrowed from non-linear dynamical systems, do not specifically account for the multi-scale nature of turbulent bursts in high-$Re$ flows. In addition to these studies, \citet{yeung2015extreme} also mention the challenging aspects associated with these bursts when the Reynolds number of the flow is increased. Particularly, \citet{yeung2015extreme} show that the topology of the structures associated with extreme events in small-scale turbulence does not necessarily scale with the increasing $Re$. In fact, their results highlight a non-trivial relationship between the large-amplitude fluctuations and the Reynolds number of a turbulent flow. Given the resurgence of interest in the topic of extreme events, it is timely to revisit this problem in high-$Re$ flows by treating the impact of multi-scale bursts on turbulence statistics through a novel framework.

\begin{figure}[h]
\centering
\hspace*{-0.6in}
\includegraphics[width=1.2\textwidth]{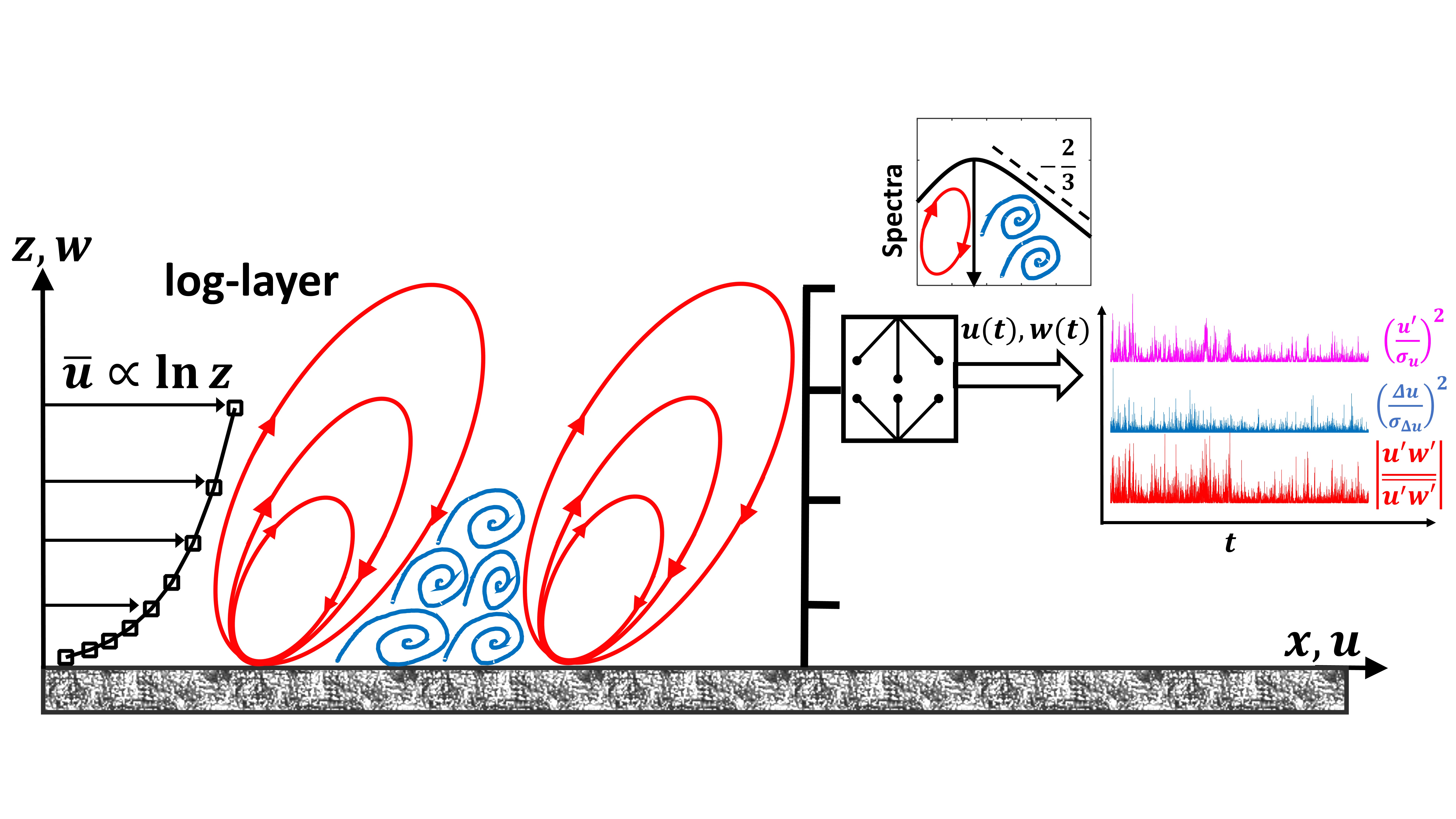}
  \caption{Schematic diagram of a near-neutral atmospheric flow is shown to indicate the presence of burst activities in various turbulence statistics. In this diagram, the $x$ axis is in the direction of the mean wind ($\overline{u}$) and the $z$ axis represents the vertical. Conceptually, the logarithmic layers (where $\overline{u} \propto \ln z$) of such flows are populated with coherent structures such as the attached eddies (shown in red) and small-scale detached eddies (shown in blue). These small-scale eddies are comparable to the inertial-subrange scales, while the scales of the coherent structures are of the order of the energetic-scale motions. This is highlighted through the premultiplied energy spectrum, which is more representative of the vertical velocity. While conducting measurements on a micro-meteorological tower, the impact of these eddy motions is registered on the instantaneous evolution of Reynolds stress components (${u^{\prime}}^2$, $u^{\prime}w^{\prime}$) and small-scale quantities such as the second-order velocity increments (${\Delta u}^2$). To illustrate this through an example, on the right-hand-side three 30-min time series of ${(u^{\prime}/\sigma_{u})}^2$ (pink), ${(\Delta {u}/\sigma_{\Delta u})}^2$ (light blue), and $\vert u^{\prime}w^{\prime}/\overline{u^{\prime}w^{\prime}}\vert$ (red) are shown from an experimental dataset (see Sect. \ref{atmosphere}).}
\label{fig:1}
\end{figure}

Before we describe the objectives of this study, it is prudent to explain how the presence of bursts affects the turbulence statistics at different scales of the flow. To illustrate this concept, in Fig. \ref{fig:1} we show a schematic of a near-neutral atmospheric surface layer flow (or equivalently, a high $Re$ wall-bounded turbulent flow). Such flows are characterized by large $Re$ values and typically occur in the lowest 10\% of the atmospheric boundary layer with a negligible effect of buoyancy on turbulence production \citep{wyngaard2010turbulence}. Moreover, in these flows, the vertical profile of the mean velocity is logarithmic ($\overline{u} \propto \ln (z)$, where $z$ is the height) and the presence of attached eddies dominates the flow statistics \citep{banerjee2013logarithmic,katul2016generalized}. However, there is also the presence of small-scale detached isotropic eddies, whose contribution is negligible to the overall flow statistics as they are sampled from the inertial subrange of the energy spectrum. This is shown through a cartoon of premultiplied spectrum in Fig. \ref{fig:1} where the inertial subrange can be identified by a $+2/3$ power-law. Nevertheless, at a given measurement level (typically on a mast), the small-scale eddies are responsible for the strong amplitude fluctuations in the velocity increments (e.g., $\Delta u(\tau)$). On the other hand, the energetic-scale eddies mainly give rise to strong amplitude variations in the bulk quantities such as the streamwise or vertical velocity fluctuations ($u^{\prime}$ or $w^{\prime}$) and instantaneous momentum flux signals ($u^{\prime}w^{\prime}$). 

Due to these differences, an interesting outcome emerges when one considers the time series of the following signals: ${u^{\prime}}^2$, ${[\Delta u(\tau)]}^2$, and $u^{\prime}w^{\prime}$. The first and last of such signals represent the time evolution of the streamwise velocity variances and momentum fluxes (Reynolds stress components), which are supposedly governed by the large-scale eddy structures. Contrarily, the middle one ${[\Delta u(\tau)]}^2$, represents the instantaneous variations in the energy content at a time scale $\tau$ of the flow ($\overline{{[\Delta u(\tau)]}^2}$). To demonstrate this point, we show an example of ${(u^{\prime}/\sigma_{u})}^2$, ${(\Delta u/\sigma_{\Delta u})}^2$, and $\vert u^{\prime}w^{\prime}/\overline{u^{\prime}w^{\prime}} \vert$ time series from a near-neutral atmospheric flow (see Fig. \ref{fig:1}). Since the momentum flux signal is a sign-definite quantity, absolute values are undertaken to better highlight their burst features. For comparison purposes, these quantities have been suitably normalized by their mean values, i.e., by variances (${\sigma^2_{u}}$, ${\sigma^2_{\Delta u}}$) and covariance ($\overline{u^{\prime}w^{\prime}}$). Notwithstanding their different origins, these three time series display qualitatively similar behavior, i.e., they all appear to be bursty (characterized by several `spikes' in the signal). However, through visual inspection, it remains a challenging task to quantify whether the turbulence generation at smaller scales of the flow is more bursty than at larger scales. In other words, answering this question requires an interlink to be established between the small- and large-scale bursts, which broadly speaking, motivates the present study. 

Conventionally, strong amplitude variations or bursts in a signal are studied through the tails of a probability density function (PDF) by employing a statistic known as kurtosis \citep{tennekes1972first,davidson2015turbulence}. The kurtosis is a fourth-order moment of any stochastic fluctuating signal $x^{\prime}$, defined as $\overline{{(x^{\prime}/\sigma_{x})}^4}$, where $\sigma_{x}$ is the standard deviation. However, since the PDF of a signal is insensitive to its temporal structure, randomly ordering the values does not have any effect on the kurtosis estimation. In this study, we revisit a quantity named "burstiness index" that can successfully account for the strong amplitude variations in a signal, while being sensitive to the signal structure. Unlike VITA or the quadrant-hole method, no arbitrary thresholds are needed for evaluating the burstiness index. Although this index had earlier been proposed by \citet{narasimha2007turbulent}, we reinterpret its physical meaning and extend its usage beyond just studying the momentum-flux signals. For instance, in contrast to previous studies, where different tools are used to investigate the small- and large-scale bursts (e.g., multifractal analysis or VITA), we adopt a novel scale-aware event-based framework to seamlessly synthesize the characteristics of small- and large-scale bursts. 

By employing this framework, we ask, (1) Do the bursts have similar physical properties when the instantaneous variations in velocity variances and momentum flux signals are considered? (2) How exactly do the burst features of such Reynolds stress components evolve as the eddy time scales in the flow increase or decrease systematically? (3) What is the role of the Reynolds number on the signal's burstiness characteristics? For assessing the Reynolds number effects, we employ datasets from two different experiments conducted in a wind tunnel and in a near-neutral atmosphere whose $Re$ values are different by almost two orders of magnitude. We restrict ourselves to near-neutral stability since at such conditions the atmospheric surface layer is known to behave analogously to a flat-plate boundary layer flow \citep{wang2016very}. The present study is organized into three different sections. In Sect. \ref{data_method}, we provide the descriptions of the experimental datasets and methodology used in this study, in Sect. \ref{results} we present and discuss the results, and lastly in Sect. \ref{conclusion} we conclude and provide future research direction.

\section{Dataset and methodology}
\label{data_method}
\subsection{Dataset}
\label{Data}
\subsubsection{Wind tunnel experiment}
\label{wind_tunnel}
One of the datasets we use is from a fully-developed turbulent boundary layer flow over an aerodynamically smooth flat plate, as obtained in the wind-tunnel facility of the University of Melbourne \citep{MARUSIC2020dataset}. The friction Reynolds number of this flow is $Re= \delta u_*/\nu \approx 14750$, where $\delta$ is the boundary-layer thickness (0.361 m), $u_{*}$ is the friction velocity (0.626 m s$^{-1}$), and $\nu$ is the kinematic viscosity of air ($1.532\times 10^{-5}$ m$^{2}$ s$^{-1}$). In this wind tunnel experiment, hot-wire anemometers were deployed to measure the time series of the streamwise velocity, $u$. The turbulent fluctuations in the streamwise velocity ($u^{\prime}$) were computed by subtracting the time-averaged mean velocity ($\overline{u}$) from $u$. These measurements were recorded at a sampling frequency ($f_s$) of $20$\,kHz for up to $120$-s at $41$ wall-normal coordinates $z$, spanning between 0.1 mm to 526 mm. Moreover, the time series of $u$ were collected for three acquisition cycles, and therefore, the results reported in Sect. \ref{results} are averaged over these three cycles. Further details of the experiment can be found in \citet{baars2015wavelet}. Throughout this study, the wall-unit normalization is indicated by the $+$ superscript such that $u^+=u/u_*$ and $z^+=zu_{*}/\nu$. Note that from the wind-tunnel experiment, only the $u^{\prime}$ signal is available and we restrict its vertical extent up to $z^{+} \leq 10^{4}$. This is because beyond that height one encounters an intermittent region where turbulent-non-turbulent patches dominate the flow behavior \citep{iacobello2021large}. 

\subsubsection{Atmospheric experiment}
\label{atmosphere}
To compare the turbulent features with an even higher Reynolds number flow, we use an atmospheric field-experimental dataset from the Surface Layer Turbulence and Environmental Science Test (SLTEST) experiment \citep{mcnaughton2007scaling,chowdhuri2019empirical}. The SLTEST experiment ran continuously for nine days from 26 May 2005 to 03 June 2005, over a flat and homogeneous terrain at the Great Salt Lake desert in Utah, USA (40.14$^\circ$ N, 113.5$^\circ$ W). The aerodynamic roughness length ($z_0$) at the SLTEST site was $z_{0}\approx$ 5 mm \citep{metzger2007near}, thereby indicating the smoothness of the surface. Although the measurement of atmospheric boundary layer depth $\delta$ was not directly available at the SLTEST site, but by assuming it around 500 m with a typical $u_{*}$ value of 0.2, the friction Reynolds number of the SLTEST experiment could be estimated as $Re=(u_{*}\delta)/\nu \approx 10^{6}$. Note that we consider $\nu= 1.8\times 10^{-5}$ m$^2$ s$^{-1}$, following \citet{marusic2013logarithmic}.

During this experiment, nine North-facing time-synchronized CSAT3 sonic anemometers were mounted on a 30-m mast, spaced logarithmically over an 18-fold range of heights, from 1.42 m to 25.7 m, with the sampling frequency ($f_s$) being set at 20 Hz. The continuous sonic anemometer data were divided into half-hour runs with each run containing the time-synchronized data from all nine sonic anemometers. In order to select the runs for our analysis, the data were subjected to various quality checks, such as stationarity, meteorological conditions at the experimental site, thresholds on the kinematic heat flux and friction velocity, satisfying the constant flux layer assumption and inertial-subrange scalings, etc. These details are outlined in \citet{chowdhuri2020representation}. 

In this study, we use a subset of 20 near-neutral runs having $-L>$ 200 m ($L$ is the Obukhov length), so that all the nine sonic anemometers lay deep within the log-layer. The friction velocity $u_{*}$ is computed as,
\begin{equation}
u_{*}={({\overline{u^{\prime}w^{\prime}}}^{2}+{\overline{v^{\prime}w^{\prime}}}^{2})}^{\frac{1}{4}},
\label{fric_velo}
\end{equation}
where $\overline{u^{\prime}w^{\prime}}$ and $\overline{v^{\prime}w^{\prime}}$ are the streamwise and cross-stream momentum fluxes respectively, at $z=$ 1.4 m. For all our selected runs, $u_*$ varied between 0.26 to 0.2. This range of $u_*$ values is in agreement with previous studies conducted in the near-neutral atmospheric surface layer \citep{haugen1971experimental}. Unless otherwise mentioned, the presented turbulence statistics in Sect. \ref{results} are ensemble-averaged over this set of near-neutral runs. While conducting the analysis on the atmospheric dataset, we focus our attention on the following signals, such as the streamwise ($u^{\prime}$) and vertical velocity fluctuations ($w^{\prime}$), and their product ($u^{\prime}w^{\prime}$), which is the instantaneous momentum flux. The turbulent fluctuations ($u^{\prime}$ and $w^{\prime}$) are computed by subtracting the 30-min linear trend from the respective variables. Henceforth, the wind tunnel and atmospheric experiments are referred to as the TBL and SLTEST experiments, respectively. In the next sub-section, we discuss the methodology to compute the burstiness index.

\subsection{Methodology}
\label{method}
 \subsubsection{Burstiness index}
\begin{figure}[h]
\centering
\hspace*{-0.9in}
\includegraphics[width=1.3\textwidth]{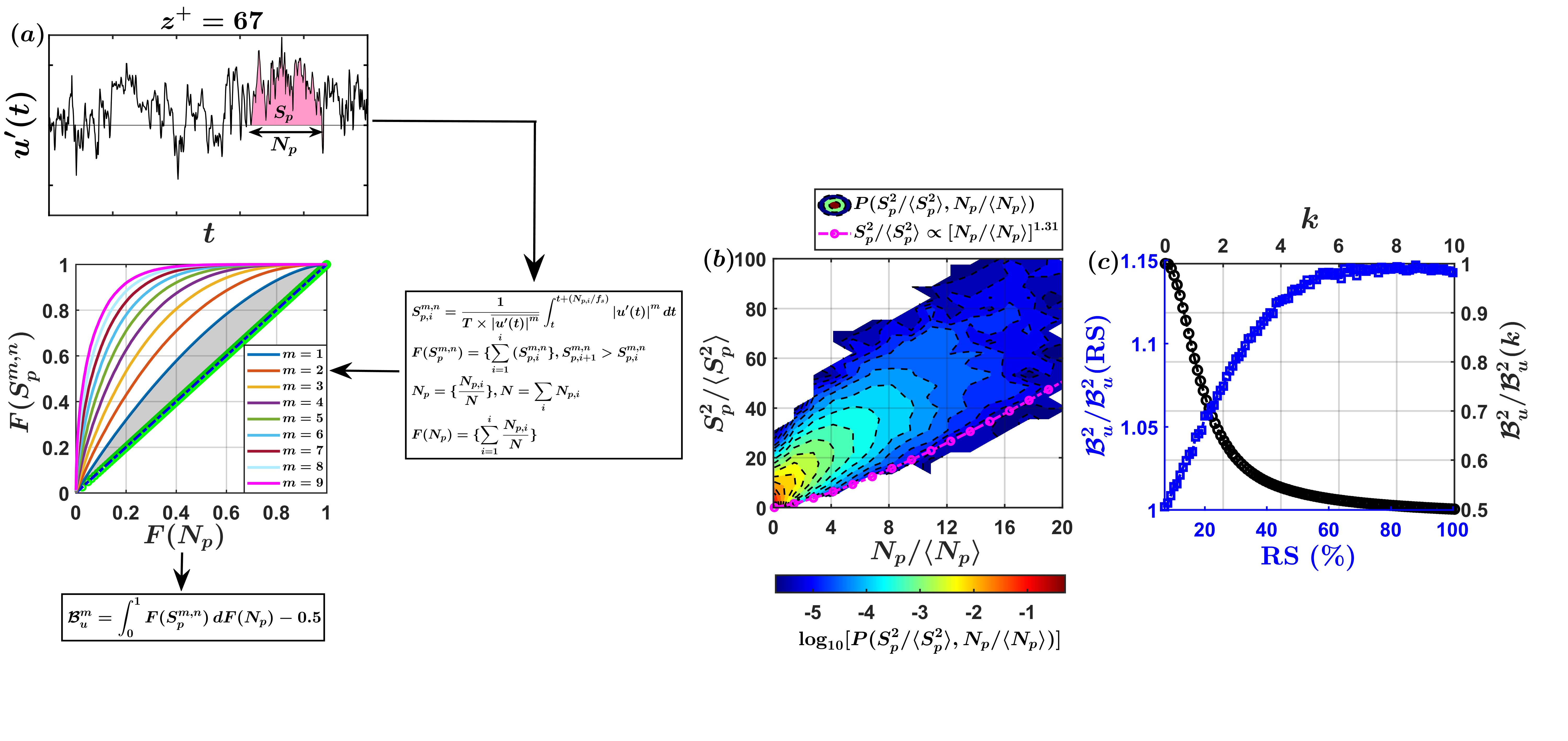}
  \caption{(a) A flowchart is shown to explain the computation of the burstiness index ($\mathcal{B}^{m}_{x}$) for any measured turbulent signal $x^{\prime}$ corresponding to its moment order $m$. For illustration purposes, we use the $u^{\prime}$ signal at $z^{+}=$ 67 from the TBL experiment. The gray-shaded region is used to highlight the area under the burstiness curve, corresponding to $m=1$. (b) For the same $u^{\prime}$ signal, the contours of the logarithms of joint probability density function (JPDF) between $N_{p}/\langle N_{p} \rangle$ and $S^{2}_{p}/\langle S^{2}_{p} \rangle$ are shown. The quantities $\langle N_{p} \rangle$ and $\langle S^{2}_{p} \rangle$ denote the averaged event length and size, respectively. The pink line with markers indicates a power-law relationship between the two. (c) The variations in $\mathcal{B}^{2}_{u}$ are shown when the $u^{\prime}$ signal is randomly shuffled (RS) in a gradual manner or when its Fourier phases are altered by changing the parameter ($k$) of a von-Mises distribution. To quantify the variations, on the left- and right-hand-side of the $y$ axis, the ratios $\mathcal{B}^{2}_{u}/\mathcal{B}^{2}_{u}(\rm RS)$ and $\mathcal{B}^{2}_{u}/\mathcal{B}^{2}_{u}(k)$ are plotted, respectively.} 
\label{fig:2}
\end{figure}

In Fig. \ref{fig:2}a, we show a section of a $u^{\prime}$ time-series from the TBL experiment at $z^{+}=$ 67. It is evident that the time-series $u^{\prime}$ undergoes transitions from positive to negative states as time evolves. Such transitions are associated with the passage of eddy structures over the measurement location \citep{cava2012role,heisel2022self}. We denote the length of any positive or negative events by $N_{p}$, which can also be transformed to a time scale $t_{p}$ after multiplying with the sampling period $1/f_s$. It is obvious that the sum over the length of all the events should be equal to the length of the time series ($N$). Corresponding to any event of length $N_{p}$, the area under the time series represents the contribution of that event to any desired turbulent statistic. For instance, if one considers the $m$-order moment of a stochastic signal, then the fractional contribution from an event (also described as event size) of length $N_{p}$ can be expressed as,
\begin{equation}
S^{m,n}_{p}=\frac{1}{T \times \overline{ {\vert u^{\prime}(t) \vert}^m}}{\int_{t}^{t+(N_{p}/f_s)} |u^{\prime}(t)|^m \,dt},
\end{equation}
where $T$ is the total duration of the time-series ($T=N/f_s$) and $\overline{ {\vert u^{\prime}(t) \vert}^m}$ is the time-averaged $m$-order moment, which for $m=2$ is simply the variance. Note that the superscript $n$ in $S^{m,n}_{p}$ is exclusively used to indicate the normalization with $\overline{ {\vert u^{\prime}(t) \vert}^m}$. Moreover, we use the absolute values of the signal while defining this quantity, such that the fractional contribution from the events to any order of the moment remains sign-indefinite and bounded between 0 to 1. It is clear that when summed over all the possible $S^{m,n}_{p}$ values it returns unity. One can also follow the same procedure for the vertical velocity signal by replacing $u^{\prime}$ with $w^{\prime}$. However, only the first-order moment is relevant for the momentum flux signal, since that represents the total time-averaged flux. In the parlance of complex systems approach, these localized event lengths and their contributions can be compared to size-duration relationships for systems exhibiting self-organized critical (SOC) behavior, such as the sandpile model for avalanche dynamics \citep{pradas2009avalanche,planet2009avalanches}. 

After defining $S^{m,n}_{p}$ and $N_{p}$, one can sort the $S^{m,n}_{p}$ values from the largest to smallest and then cumulatively sum them together. This cumulative sum converges to unity, since $S^{m,n}_{p}$ values are divided by $\overline{ {\vert u^{\prime}(t) \vert}^m}$. Let us denote this cumulative sum as $F(S^{m,n}_{p})$. Similarly, corresponding to the sorted values of $S^{m,n}_{p}$, one can cumulatively sum the event lengths by normalizing them with respect to the length of the time series ($N$). We denote this as, $F(N_{p})$. As a next step, $F(S^{m,n}_{p})$ and $F(N_{p})$ are plotted against one another, which one refers to as a burstiness curve. An example of such a curve is shown in Fig. \ref{fig:2}a, where different moment orders are plotted ($m=$ 1 to 9). 

We next explain how such a plot between $F(S^{m,n}_{p})$ and $F(N_{p})$ can be used to infer the strength of the amplitude variations, thereby capturing the effect of the turbulent bursts. If one considers a signal without any amplitude variation but only the lengths of the positive and negative events are preserved (otherwise known as a telegraphic approximation (TA)), then for such a signal the burstiness curve would be a straight line with a slope of $45^{\circ}$. This is because the fractional contributions of the events will be identical to the length up to which the events persist. We illustrate this by creating a synthetic signal of $u^{\prime}(t)$ whose all values are replaced with $\pm \sigma_{u}$, where the sign depends on the original signal. Thereafter, if we plot $F_{\rm TA}(S^{m,n}_{p})$ against $F_{\rm TA}(N_{p})$, then, as expected, the points fall exactly on the $45^{\circ}$ line (shown as green circles on the burstiness curve in Fig. \ref{fig:2}a).  

Therefore, the further the plot between $F(S^{m,n}_{p})$ and $F(N_{p})$ differ from the straight line (representing $F_{\rm TA}(S^{m,n}_{p})$ vs. $F_{\rm TA}(N_{p})$), stronger amplitude variations are present in the signal, and hence, they appear more bursty. This is reflected in Fig. \ref{fig:2}a, where one observes if the moment orders are increased (thereby enhancing the importance of the extreme events), the curves significantly deviate from the straight line. One can thus use the area under the curve between $F(S^{m,n}_{p})$ and $F(N_{p})$ and subtract it from 0.5 (which is the area under the $45^{\circ}$ straight line) to quantify the peaked nature of a signal. For illustration purposes, in Fig. \ref{fig:2}a, we shade this area in grey for the burstiness curve corresponding to $m=1$. To numerically compute the area under the burstiness curve, we use a trapezoidal approximation. This area with 0.5 subtracted is referred to as a burstiness index and denoted by $\mathcal{B}^m_{x}$, where $m$ is the moment order and $x$ is the signal under investigation. 

This whole procedure behind the computation of the burstiness index is graphically illustrated through a flow chart in Fig. \ref{fig:2}a. The burstiness index will be 0 if no amplitude variation is present in the signal. On the other hand, the maximum value of a burstiness index will be 0.5, because both $F(S^{m,n}_{p})$ and $F(N_{p})$ are bounded between 0 to 1, and therefore, the burstiness curve cannot cross the upper half of the triangle. Further utilities of the burstiness index are explained below.

Out of all the moment orders, one particular quantity of interest is the ${u^{\prime}}^2(t)$ signal, since it represents the instantaneous variations in the streamwise velocity variance. To explore the temporal evolution of ${u^{\prime}}^2(t)$, one can investigate the joint probability density function (JPDF) between $S^{2}_{p}$ and $N_{p}$. Note that $S^{2}_{p}$ is the unscaled version of $S^{2,n}_{p}$ that encapsulates the amplitude information, and hence, would depend on the signal PDF. A similar approach was taken by \citet{planet2009avalanches} while analyzing the complex interfacial dynamics of the imbibition fronts. They mentioned the quantities $S_{p}$ and $N_{p}$ as avalanche sizes and lengths, respectively, and normalized them by their mean values $\langle S_{p} \rangle$ and $\langle N_{p} \rangle$. Mathematically, these mean quantities are defined as, 
\begin{equation}
\langle x_{p} \rangle=\frac{1}{\mathcal{Z}}\sum_{i=1}^{\mathcal{Z}} x_{p,i}, \ x=\{N,S\},
\end{equation}
where $\mathcal{Z}$ is the number of zero-crossings in the signal. \citet{planet2009avalanches} found the JPDF between $S_{p}/\langle S_{p} \rangle$ and $N_{p}/\langle N_{p} \rangle$ followed a power-law variation with a slope of 1.31, which they attributed to the presence of burst-like activities in the interfacial dynamics. In agreement with \citet{planet2009avalanches}, we observe the JPDFs between $S^{2}_{p}/\langle S^{2}_{p} \rangle$ and $N_{p}/\langle N_{p} \rangle$ follow a power-law scaling for the ${u^{\prime}}^2(t)$ signal at $z^{+}=$ 67 (Fig. \ref{fig:2}b). For comparison purposes, we show the same power-law of \citet{planet2009avalanches} as a pink line with markers in Fig. \ref{fig:2}b. Therefore, the temporal evolution of the instantaneous streamwise velocity variance exhibits a complex structure, and through Fig. \ref{fig:2}c, we show the burstiness index of ${u^{\prime}}^2(t)$ can indeed capture such features. 

Since the event contributions to variance and their lengths are strongly interlinked (as seen through their JPDFs in Fig. \ref{fig:2}b), the burst-like features of a signal should depend on both PDFs of the signal and event duration. To disentangle these aspects, we employed two different surrogate signals. One of the surrogate signals was generated through gradual random shuffling. In this method, the signal PDFs are preserved but the PDFs of event lengths approach a Poisson distribution as the strength of the random shuffling (RS) is increased. The second surrogate signal exploits the Fourier phase-alteration technique (see Appendix \ref{app_A}), through which we preserve the PDFs of event lengths but introduce more extreme events in the signal, thereby affecting its PDF. The alteration of the Fourier phases is achieved through a von-Mises parameter $k$. As demonstrated in the appendix (see Appendix \ref{app_A}), further the parameter $k$ deviates from zero more large-amplitude spikes appear in the signal. Notice that for both such surrogate signals, the variance remains the same as the original one. More details on these surrogate data generation techniques can be found in Appendix \ref{app_A}.

In Fig. \ref{fig:2}c, we plot the ratios of the burstiness indices between the original and randomly-shuffled ($B^{2}_{u}/B^{2}_{u}(\rm RS)$) or phase-altered ($B^{2}_{u}/B^{2}_{u}(k)$) signals. One can see that as the strength of the randomization increases (i.e., the temporal coherence is gradually destroyed), $B^{2}_{u}(\rm RS)$ decreases which implies the burstiness index is dependent on the temporal structure of the signal. On the contrary, as the extreme events in the signal increase (by increasing $k$) but maintain the temporal coherence through event length PDFs, $B^{2}_{u}(k)$ attain larger values. By combining the two, one can infer the burstiness index explains the strong amplitude fluctuations in a signal by taking into account both the signal's complex structure and its PDF. In the next section, we show how a similar approach can be adopted to evaluate the scale dependence of the burstiness index.

\subsubsection{Scale-dependence of the burstiness index}

\begin{figure}[h]
\centering
\hspace*{-0.8in}
\includegraphics[width=1.6\textwidth]{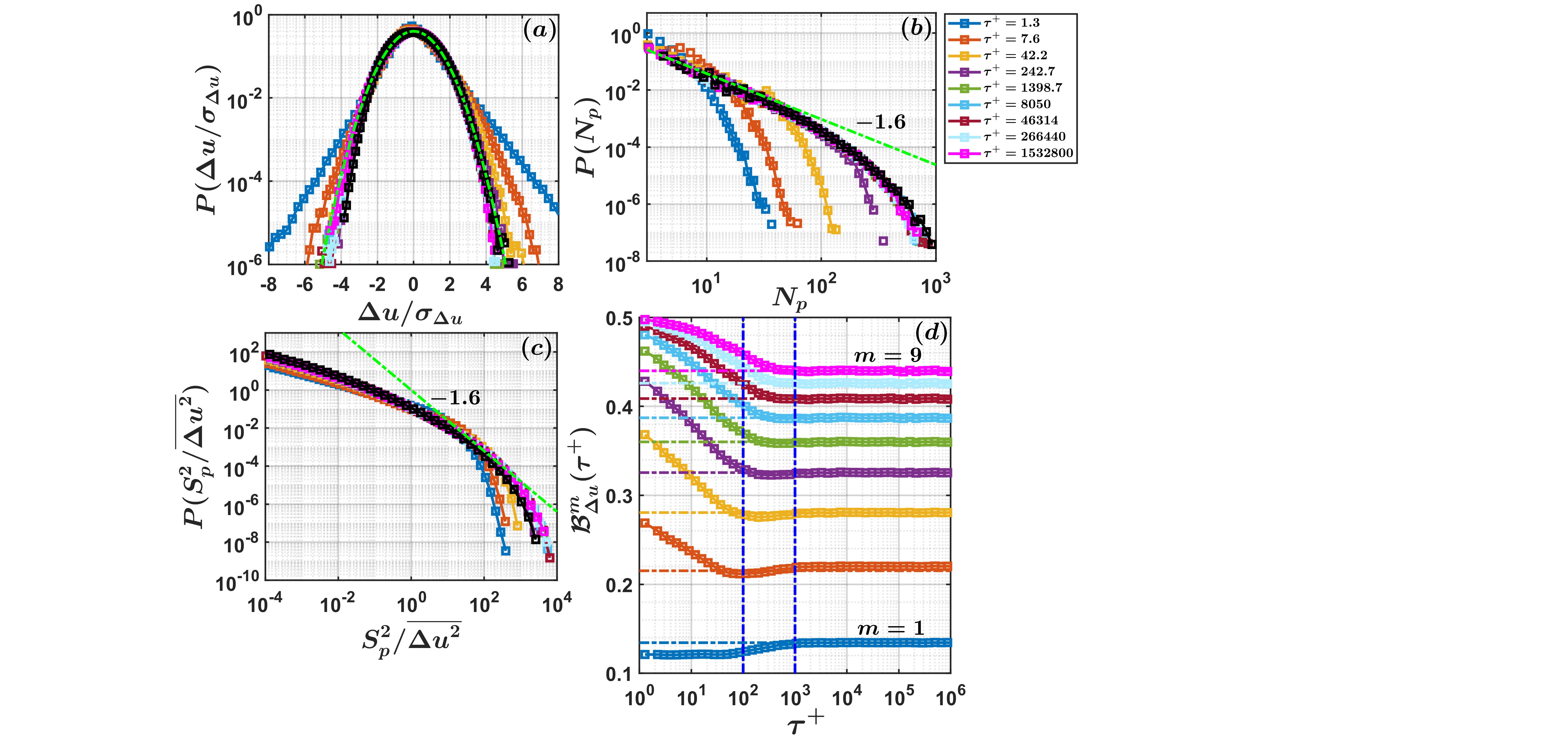}
  \caption{(a) The probability density functions (PDF) of normalized velocity increments ($\Delta u/\sigma_{\Delta u}$) are plotted for various time lags (see the legend), corresponding to the $u^{\prime}$ signal at $z^{+}=$ 67. The time lags are normalized by the inner-scaling ($\tau^{+}$) and the black line indicates the PDF of $u^{\prime}/\sigma_{u}$. The green dash-dotted line represents the Gaussian distribution. (b) The PDFs of event lengths ($N_{p}$) are shown for the velocity increment ($\Delta u(\tau^{+})$) signals at prescribed $\tau^{+}$ values. The black line indicates the PDF of $N_{p}$ computed for the $u^{\prime}$ signal. A $-1.6$ power law is shown by the green dash-dotted line. (c) For the same $\tau^{+}$ values, the PDFs of event sizes normalized by the variances of the velocity increments ($\overline{{\Delta u}^2}$) are shown. The black line indicates the event size PDFs of the $u^{\prime}$ signal. In (d) we illustrate the scale-dependence of the burstiness index ($\mathcal{B}^{m}_{\Delta u}(\tau^{+})$), as evaluated for the velocity increments ($\Delta u$) and their moment order $m$. The moment orders increase as one progresses from the light-blue color ($m=1$) to the pink one ($m=9$). The two vertical lines in (d) denote the inner- and outer-spectral peak positions from the TBL experiment. The horizontal lines indicate the values of $\mathcal{B}^{m}_{u}$.}
\label{fig:3}
\end{figure}

One of the intriguing results in fully-developed turbulent flows is the velocity increments (for example, $\Delta u(\tau)$) are increasingly non-Gaussian as the time-lags ($\tau$) are reduced \citep{sreenivasan1997phenomenology}. Therefore, with decreasing $\tau$, the importance of extreme amplitude variations becomes more evident. Instead of studying this phenomenon through just the PDFs of $\Delta u(\tau)$, one can extend the event framework to the velocity increments at any prescribed time lag and compute its burstiness index. For instance, at a time-lag $\tau$, one can define event sizes and lengths analogous to Fig. \ref{fig:2}a by considering $\Delta u(\tau)$ as the relevant signal. We illustrate this through an example in Fig. \ref{fig:3}. Henceforth, the normalized time-lags with respect to the wall-unit scaling are denoted as $\tau^{+}$.

From Fig. \ref{fig:3}a, one can clearly see as $\tau^{+}$ decreases (see the legend for different colors) the normalized PDFs of velocity increments ($P(\Delta u/\sigma_{\Delta u})$, where $\sigma_{\Delta u}$ is the standard deviation of $\Delta u$ at any given lag) become significantly non-Gaussian. If one compares the distributions of event lengths for those lags, it can be noticed that at the smallest $\tau^{+}$ value $P(N_{p})$ decreases quite rapidly (Fig. \ref{fig:3}b). However, as $\tau^{+}$ increases, $P(N_{p})$ gradually approach the event-length PDFs as obtained from the $u^{\prime}$ signal (solid black line) --- having a distinct power-law section with an exponent $-1.6$ (Fig. \ref{fig:3}b). This implies the event-length PDFs of the $u^{\prime}$ signal encompass the cumulative effects of all the flow structures passing over the measurement location. On the other hand, if the PDFs of event contributions $P(S_p^2)$ (or event sizes) to the variances for the velocity increment $\Delta u$ signals are considered at any $\tau^{+}$ values and compared with the result obtained from the $u^{\prime}$ signal, no such clear dependence on $\tau^{+}$ can be noted (Fig. \ref{fig:3}c). Therefore, the event features of the $\Delta u$ signal evolve in a non-trivial fashion as $\tau^{+}$ increases.

To explore this further, one can study the burstiness curves at any prescribed time-lag. In Fig. \ref{fig:3}d, we show the scale-dependent burstiness indices ($\mathcal{B}^m_{\Delta u}(\tau^{+})$) of the signal $\vert {\Delta u}^m(\tau^{+}) \vert$, corresponding to its moments ($m$) of the order 1 to 9. We consider the absolute values of velocity increments, which is regarded as a standard practice in turbulence literature while conducting structure-function analysis \citep{benzi1993extended}. In Fig. \ref{fig:3}d, $m$ progressively increases from the light-blue ($m=1$) to pink ($m=9$) color. The dash-dotted horizontal lines of the same color as the curves indicate the $\mathcal{B}^{m}_{u}$ values. One can notice that, except for $m=1$, the burstiness indices vary similarly for any other $m$ values. For instance, $\mathcal{B}^2_{\Delta u}(\tau^{+})$ attains a maximum at the smallest possible $\tau^{+}$ and then decreases with increasing lags. Eventually, they saturate to the values ($\mathcal{B}^{m}_{u}$) as obtained from the full-signal ${[u^{\prime}(t)]}^m$. More importantly, such saturation typically occurs at scales commensurate with the outer spectral peak at $\tau^{+}=1000$ \citep{baars2015wavelet}. Therefore, this outcome points towards a seamless transition from small- to large-scale bursts as the eddy time scales increase. Note that the inner- ($\tau^{+}=100$) and outer-spectral ($\tau^{+}=1000$) peak positions are estimated from the premultiplied $u$ spectra presented in \citet{baars2015wavelet}. 

The saturation to the full-signal values ($\mathcal{B}^{m}_{u}$) indicate that the large-scale structures mainly govern the burst features observed in the ${[u^{\prime}(t)]}^m$ signals. On the other hand, strong amplitude variations in velocity increments are mainly confined to the small-scale motions. Although not shown here, but through synthetic turbulence data one can ascertain that the behavior of $\mathcal{B}^m_{\Delta u}(\tau^{+})$ with increasing lags is sensitive to the multifractal nature of small-scale turbulence \citep{sreenivasan1991fractals}. 

Hereafter, we will focus on the second- and mixed-order velocity increments, such as, ${\Delta u}^2(\tau^{+})$, ${\Delta w}^2(\tau^{+})$, and $\Delta u \Delta w(\tau^{+})$. As an alternative to Fourier spectrum or cospectrum, the averages of these quantities (e.g., $\overline{{\Delta u}^2(\tau^{+})}$) physically represent the contribution to Reynolds stress components (e.g., $\sigma_{u}^2$) at any specified scale of the flow \citep{schmitt2016stochastic}. Hence, the variations in $\mathcal{B}^2_{\Delta x}(\tau^{+})$ ($x=u,w$) and $\mathcal{B}^1_{\Delta u \Delta w}(\tau^{+})$  with increasing time-lags would quantify the role of bursts on the scale-wise evolution of Reynolds stress components. Since $\Delta u \Delta w$ is a sign-definite quantity, we use their absolute values ($|\Delta u \Delta w|$) while computing $\mathcal{B}^1_{\Delta u \Delta w}(\tau^{+})$. 

\subsubsection{Randomly-shuffled and IAAFT signals}
To underpin what flow features are responsible behind the turbulent bursts, we use two different surrogate signals. One of them is generated through a random-shuffling procedure. In this method, a random permutation is operated on a time-series to disrupt the underlying temporal arrangement, thereby creating a surrogate dataset that does not possess any relationship among the signal data points. Therefore, in randomly-shuffled surrogates, the signal's PDF remains precisely conserved albeit the data points appear random. 

The second type of surrogate is generated from a procedure named as iteratively adjusted amplitude Fourier transform (IAAFT). The IAAFT surrogates do not contain non-linear effects but preserve the linear effects described by the auto-correlation or Fourier spectrum of the time series \citep{lancaster2018surrogate}. This is accomplished by keeping the Fourier amplitudes of the time series intact, but replacing the associated Fourier phases with a random uniform distribution between 0 to 2$\pi$. The randomness in the Fourier phases destroys any non-linear structure of the time series. However, due to the randomisation of the Fourier phases the PDF of the time series becomes Gaussian. Hence, to preserve both PDF and amplitude spectrum, the Fourier amplitudes and the signal's PDFs are adjusted iteratively at each stage of phase randomisation until the resultant signal has the same power spectrum and the PDF as the original one. 

In the context of turbulent signals, if the results from an IAAFT surrogate signal are compared with a randomly-shuffled one, then the difference between the two can be directly associated with the energy spectrum. Therefore, this comparison enables one to ascertain the effect of coherent structures (which contribute the most to the turbulence kinetic energy) on the desired turbulent statistic.

\section{Results and discussion}
\label{results}
We begin with comparing the turbulence statistics between the TBL and SLTEST datasets. Such comparisons enable us to infer the type of coherent structures present in both flows. Thereafter, we focus on the scaling properties of the event time scales and their magnitudes to probe the effects of the flow structures on the peaked nature of velocity and momentum flux signals. To the best of our knowledge, this is the first time event-based features are compared between the laboratory and atmospheric flow settings. Furthermore, we introduce a novel scale-dependent event framework through which we establish a statistical correspondence between the event and eddy time scales. We conclude our study by applying this framework to quantify the effect of turbulent bursts on velocity variances and momentum transport at each scale of the flow. 

\subsection{Comparison between the laboratory and atmospheric flows}
\label{comparison}

\begin{figure}[h]
\centering
\hspace*{-0.8in}
\includegraphics[width=1.3\textwidth]{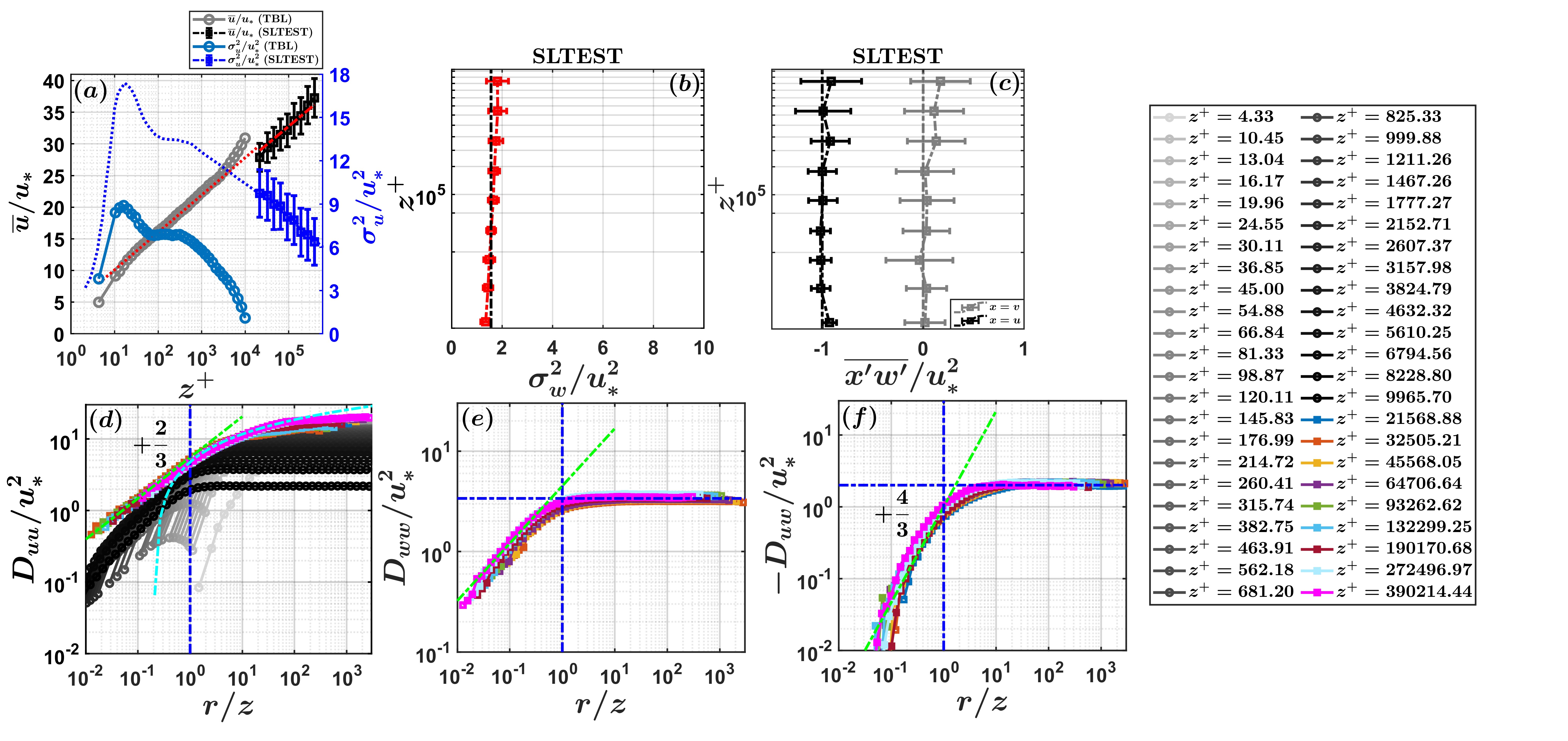}
  \caption{(a) The vertical profiles of the normalized mean velocity ($\overline{u}/u_{*}$) and variances ($\sigma_{u}^2/u^{2}_{*}$) are compared between the TBL and SLTEST datasets (see the legend). The blue dotted line in (a) is digitized from Fig. 1a of \citet{yang2018scaling}. The red lines denote the logarithmic fits of \citet{marusic2013logarithmic} to the mean velocity profile. The profiles of (b) normalized vertical velocity variances ($\sigma_{w}^2/u^{2}_{*}$) and (c) streamwise and cross-stream momentum fluxes ($\overline{x^{\prime}w^{\prime}}/u_*^2$, where $x$ can be $u$ or $v$) are presented from the SLTEST dataset. In (d)--(f), normalized  second-order structure functions $D_{uu}/u_*^2$, $D_{ww}/u_*^2$, and mixed-order structure function $-D_{uw}/u_*^2$ are plotted against $r/z$, where $r$ is the spatial lag and $z$ is the height. The green dash-dotted lines in (d)--(f) indicate the inertial-subrange slopes of $+2/3$ and $+4/3$, respectively. The cyan-colored line in (d) denotes the logarithmic scaling of $D_{uu}/u_*^2$ at larger scales of the flow. The legend at the extreme left end represents the color codes corresponding to the heights from the TBL and SLTEST experiments.}
\label{fig:4}
\end{figure}

\subsubsection{Turbulence statistics}
Figure \ref{fig:4}a--c show the vertical profiles of the mean velocity ($\overline{u}/u_{*}$), velocity variances ($\sigma_u^2/u_{*}^2$ and $\sigma_w^2/u_{*}^2$), and streamwise and cross-stream momentum fluxes ($\overline{u^{\prime}w^{\prime}}/u_{*}^2$ and $\overline{v^{\prime}w^{\prime}}/u_{*}^2$). These quantities and the height ($z$) are normalized with the wall-unit scaling, such as by $u_{*}$ and $\nu$. The error bars denote the spread from the ensemble mean for the SLTEST dataset. 

From Fig. \ref{fig:4}a, one can notice that in the TBL experiment, the mean velocity profile stays logarithmic up to a certain height range (red dotted line). Accordingly, the SLTEST dataset too maintains a logarithmic mean velocity profile (red dash-dotted line). The curves to fit the logarithmic variations are adopted from \citet{marusic2013logarithmic}. As per Townsend's attached eddy hypothesis \citep{townsend1980structure}, the streamwise velocity variances are supposed to follow a logarithmic scaling in the inertial layer of a wall-bounded turbulent flow \citep{banerjee2013logarithmic}. However, such scaling involves the outer-layer variables (boundary-layer height, $\delta$), and therefore, cannot be directly compared between the two experiments. Despite this limitation, vertical profiles of streamwise velocity variances are characteristically similar between the TBL and SLTEST experiments. This is illustrated through the blue dotted line in Fig. \ref{fig:4}a. The blue dotted line is digitized from Fig. 1a of \citet{yang2018scaling}. In that figure, \citet{yang2018scaling} adopt a semi-empirical formulation of $\sigma_u^2/u_{*}^2$-profile from \citet{kunkel2006study} to fit their near-neutral atmospheric dataset. Our observations indicate that the streamwise velocity variances of the SLTEST experiment match nicely with this prediction.

In contrast to the streamwise velocity variances, $\sigma_w^2/u_{*}^2$ (Fig. \ref{fig:4}b) remain constant with height, with the constant being equal to the square of 1.25, as empirically observed by \citet{kader1990mean}. On the other hand, the normalized streamwise momentum fluxes ($\overline{u^{\prime}w^{\prime}}$) remain equal to the friction velocity value at the surface, while the cross-stream component ($\overline{v^{\prime}w^{\prime}}$) is nearly 0 (Fig. \ref{fig:4}c). This indicates the surface shear stress aligns with the direction of the mean wind \citep{bernardes2010alignment}. 

\subsubsection{Structure function analysis}
All such bulk statistics are in confirmation with Townsend's attached eddy model, and hence, the coherent structures present in both flows are supposedly the attached eddies. It is therefore expected that the impact of such attached eddies would reflect in the behavior of the energy spectrum or second-order structure functions. Here we focus on the structure functions ($\overline{{\Delta u(\tau)}^2}$) since these statistics are later used while investigating the scale-wise behavior of turbulent bursts (Figs. \ref{fig:6}--\ref{fig:7}). Note that the $u$ spectra from the TBL dataset are presented in \citet{baars2015wavelet} and regarding the SLTEST dataset, $u$, $w$ spectra, and $u$-$w$ co-spectra are shown in Appendix \ref{app_B} (Fig. \ref{fig:9}). In all the following figures (Figs. \ref{fig:4}--\ref{fig:10}), two different color schemes are mostly used to demarcate between the TBL and SLTEST experiments. For instance, grey-shaded lines with varying intensities represent the TBL dataset while the colored lines are from the SLTEST experiment (see the legend of Fig. \ref{fig:4}). Specific to the TBL dataset, the faintest color indicates the lowest height ($z^{+}=4.33$) and the darkest one corresponds to $z^{+}=9965.70$.   

In Fig. \ref{fig:4}d, we compare the scaling behavior of the streamwise velocity structure functions ($D_{uu}/u_{*}^2$, where $D_{uu}=\overline{\Delta u(\tau)^2}$) between the TBL and SLTEST experiments. As commonly done while studying the scaling properties of structure functions, we convert the time-lags ($\tau$) to spatial lags ($r=\tau \overline{u}$) by using the Taylor's hypothesis \citep{dixit2013scaling,cheng2020power}. Regardless of TBL or SLTEST datasets, $\sigma_{u}/\overline{u}$ was less than 0.5, thereby affirming the validity of the Taylor's hypothesis \citep{willis1976use}. For both experiments, one can notice that at scales comparable to the inertial-subrange ($r<z$), $D_{uu}/u_{*}^2$ follow the $+2/3$ Kolmogorov scaling. On the other hand, a log-scaling is observed at the energetic scales ($r>z$) of motion. The cyan colored line in Fig. \ref{fig:4}d shows the fitted log-scaling as adopted from \citet{ghannam2018scaling}.

This log-scaling is expressed as, $D_{uu}/u_{*}^2=A\ln(r/z)+B$, where $A$ and $B$ are 2.5 and 1.8, respectively \citep{ghannam2018scaling}. Physically, the presence of log-scaling in the structure functions is a tell-tale sign of attached eddies in the flow, reflected as a $\kappa^{-1}$ scaling ($\kappa$ is the wavenumber) in the $u$ spectrum \citep{perry1977asymptotic,perry1987experimental,banerjee2013logarithmic}. Interestingly, for the SLTEST dataset, the attached-eddy scaling is more prominent in $D_{uu}/u_{*}^2$ rather than in its spectral counterpart (Fig. \ref{fig:9}). Moreover, in accordance with the attached-eddy model, such log-scaling is absent in $D_{ww}/u_{*}^2$, although its $+2/3$ slope remains intact (Fig. \ref{fig:2}e). In particular, $D_{ww}/u_{*}^2$ approach $2{\sigma_{w}}^2/{u_{*}}^2$ as the scales increase (horizontal blue dashed line in Fig. \ref{fig:4}e). However, in agreement with \citep{chamecki2004local}, the spectral ratio $D_{ww}/D_{uu}$ remains smaller than the isotropic prediction of 4/3 in the inertial subrange scales (not shown).

Regarding $u^{\prime}w^{\prime}$, similar to $u$-$w$ cospectra, mixed-order structure functions $-D_{uw}/u_{*}^2$ ($D_{uw}=\overline{\Delta u \Delta w}$) describe the scale-dependent features of momentum transport \citep{mydlarski2003mixed,chamecki2017scaling}. The negative sign in $D_{uw}$ is to ensure that the quantity stays positive. At the inertial-subrange scales, $-D_{uw}/u_{*}^2$ are observed to follow the $+4/3$ scaling as per \citet{Wyn72} (Fig. \ref{fig:4}f). However, at energy-production scales ($r>z$), $-D_{uw}/u_{*}^2$ attain a constant value of 2 (horizontal blue dashed line in Fig. \ref{fig:4}f). This indicates almost all the momentum transport are accomplished through such scales. More precisely, at energy-production scales, the ejection and sweep motions emerge as the major transporters of streamwise momentum flux (see Appendix \ref{app_B}). Previous studies have shown that these ejection and sweep structures are ultimately connected to the attached eddies in the flow \citep{hommema2003packet}. 

As a side-note, since the computation of burstiness index of momentum flux signals involve absolute values, it is imperative to evaluate how the scaling behavior changes if instead of $\overline{\Delta u \Delta w}$, $\overline{|\Delta u \Delta w|}$ is used. Due to its absolute nature, we find that the overall scale-wise evolution of $\overline{|\Delta u \Delta w|}/u_{*}^2$ remains similar but the slope of inertial-subrange empirically changes from $+4/3$ to $+1/2$ (see Fig. S1 in supplementary material). Be that as it may, after establishing the fact that both TBL and SLTEST flows are in sync with the attached-eddy picture, we move to the next section where we explore how the presence of such eddy structures are reflected in the statistics of event sizes and duration. 

\subsection{Event characteristics of laboratory and atmospheric flows}
\label{Burstiness}

\begin{figure}[h]
\centering
\hspace*{-0.5in}
\includegraphics[width=1.2\textwidth]{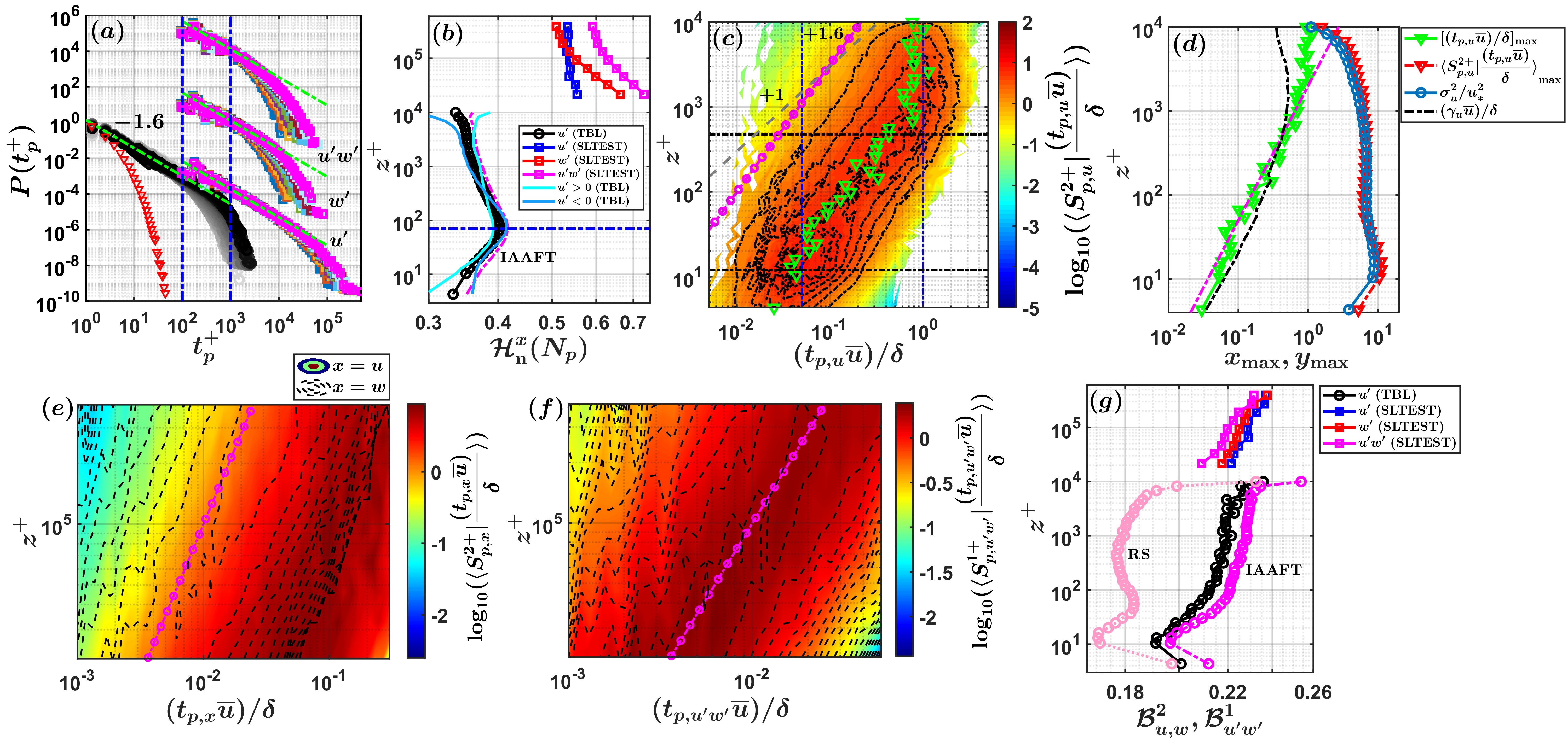}
  \caption{(a) The PDFs of normalized event time scales ($P(t_{p}^{+})$) are shown corresponding to the $u^{\prime}$, $w^{\prime}$, and $u^{\prime}w^{\prime}$ signals. The time scale PDFs of $w^{\prime}$, and $u^{\prime}w^{\prime}$ signals are shifted vertically upwards. The green dash-dotted lines show a power-law scaling with an exponent of $-1.6$. (b) The vertical profiles of normalized entropy of event lengths ($\mathcal{H}^{x}_{n}(N_{p})$) are shown. For the TBL flow, $\mathcal{H}^{x}_{n}(N_{p})$ is compared with an IAAFT surrogate signal (pink dashed line), and with $u^{\prime}>0$ (cyan solid line) and $u^{\prime}<0$ events (light-blue solid line). The dash-dotted blue horizontal line indicates $z^{+}=$ 70. (c) The contours of event contribution curves are plotted against the normalized event length scales ($t_{p}\overline{u}/\delta$) and $z^{+}$, corresponding to the TBL experiment. The green markers show those $t_{p}\overline{u}/\delta$ which contribute the most to the velocity variance. The grey dashed line and pink line with circles denote $+1$ and $+1.6$ power-laws, respectively. (d) The vertical profiles of the maxima of event-contribution curves are shown from the TBL experiment. These maxima are compared with the normalized integral scales of $u^{\prime}$ ($(\gamma_{u}\overline{u})/\delta$) and velocity variances ($\sigma_{u}^2/u_{*}^2$). In (e) and (f), the contours of event contribution curves towards the velocity variances and momentum fluxes are shown from the SLTEST experiment. (g) The vertical profiles of burstiness index ($\mathcal{B}^{2}_{u,w}$, $\mathcal{B}^{1}_{u^{\prime}w^{\prime}}$) are shown for the SLTEST and TBL experiments. This index is compared with randomly-shuffled (RS) and IAAFT surrogate signals shown as light- and dark-pink lines, respectively.}
\label{fig:5}
\end{figure}

\subsubsection{Event time scales}
Figure \ref{fig:5}a shows the PDFs of event time scales ($t_{p}=N_{p}/f_s$), corresponding to $u^{\prime}$, $w^{\prime}$, and $u^{\prime}w^{\prime}$ signals. The event time scales are normalized in wall units ($t_{p}^{+}$) so that the vertical variations can be identified in $P(t_{p}^{+})$. The computation procedure of these PDFs is similar to as described in \citet{chowdhuri2020persistence}. It can be proven that these PDFs encode the effect of the turbulent structures in the flow. For instance, if one randomly shuffles the turbulent signal (thereby destroying all the ordered structures) and recomputes these PDFs, the result is very different from the original (shown as red triangles in Fig. \ref{fig:5}a). For comparison purposes, $P(t_{p})$ of a randomly shuffled signal is an exponential distribution and has a kurtosis of 9 \citep{weber2019wind}. However, the kurtosis of original event time scales ($\mathcal{K}(t_{p})$) exceed 9 considerably and can attain values as large as 100 (see Fig. S2c in supplementary material). 

For the $u^{\prime}$ signals from the TBL experiment, one observes a power-law segment with an exponent of $-1.6$ in $P(t_{p}^{+})$. This power-law segment extends almost up to the time scales commensurate with the outer-spectral peak position ($t_{p}^{+}=$ 1000). Beyond that, the PDFs deviate from the power-law behavior and a clear height variation is observed, implying that the larger time-scale events become more probable as the heights increase. On the contrary, for the same signals from the SLTEST experiment, one notices hardly any difference among different heights. Nevertheless, the same power law is found to be present for the SLTEST data too, despite their extent being different. It can be shown that under a different scaling (for instance, using $\delta$ as a scaling height), $P(t_{p})$ between the TBL and SLTEST experiments compare quite nicely (see Fig. S3a in the supplementary material). For the $w^{\prime}$ and $u^{\prime}w^{\prime}$ signals, at larger $t_{p}^{+}$ values, a height dependence is observed in the PDFs. These PDFs can be successfully collapsed if $z$ is used as a scaling parameter, thereby affirming the existence of the locally attached eddies (not shown).

To quantify these different behaviors of $P(t_{p}^{+})$, one can compute its Shannon entropy. Moreover, to relate this quantity to the organization of the flow structures, the Shannon entropy ($\mathcal{H}$) is normalized with respect to a randomly shuffled (RS) signal. However, for accuracy purposes, it is recommended to use the event lengths ($N_{p}$) instead of their time scales ($t_{p}=N_{p}/f_s$). We do this because the set of event lengths are natural numbers (e.g., $\{1,2,3,\dots\}$) rather than a continuous variable. Therefore, their PDFs ($P(N_{p})$) transform to probability mass functions (PMFs) whose computation does not suffer from any arbitrary binning \citep{paninski2003estimation}. As a result, normalized Shannon entropy of the event lengths ($N_{p}$) are defined as,
\begin{equation}
\mathcal{H}^{x}_{n}(N_{p})=\frac{\sum_{i}P(N^{x,r}_{p,i})\log[P(N^{x,r}_{p,i})]}{\sum_{i}P(N^{x}_{p,i})\log[P(N^{x}_{p,i})]},
\label{SN}
\end{equation}
where $x$ is the signal under investigation ($x=u^{\prime}, w^{\prime}, u^{\prime}w^{\prime}$), $N^{x,r}_{p}$ denotes the event lengths from a RS sequence of $x$, and $P(N^{x,r}_{p})$ are their associated probabilities. Note that $\mathcal{H}^{x}_{n}(N_{p})$ is bounded between 0 to 1, as the entropy is maximum for an RS sequence. Since an RS sequence is devoid of any order, further the deviation of $\mathcal{H}^{x}_{n}(N_{p})$ from 1, more organized the flow is. In Fig. \ref{fig:5}b, we show the vertical profiles of $\mathcal{H}^{x}_{n}(N_{p})$ associated with $u^{\prime}$, $w^{\prime}$, and $u^{\prime}w^{\prime}$ signals.

From Fig. \ref{fig:5}b, one notices the profiles of the Shannon entropies are different between the two flows. The $\mathcal{H}^{u^{\prime}}_{n}(N_{p})$ of the TBL experiment remains significantly lower than its counterpart from the SLTEST experiment, thereby indicating more organization. Specific to the SLTEST dataset, $\mathcal{H}^{u^{\prime}}_{n}(N_{p})$ values are nearly constant with height. On the other hand, $\mathcal{H}^{w^{\prime}}_{n}(N_{p})$ and $\mathcal{H}^{u^{\prime}w^{\prime}}_{n}(N_{p})$ increase with height, albeit at different rates. Some recent works have indeed pointed out that although a hierarchy of attached eddies supposedly governs the flows in a neutral atmospheric surface layer and in a laboratory setting, their organization is not similar and depends on the flow configuration \citep{liu2021large}. Interestingly, such conclusions in the previous studies have been drawn from a spectral perspective, but our results demonstrate for the first time that even from an event perspective the same principle holds. 

Moreover, $\mathcal{H}^{u^{\prime}}_{n}(N_{p})$ of the TBL experiment shows a clear inflection in its vertical profile at around $z^{+}=70$ (denoted as a blue dash-dotted horizontal line in Fig. \ref{fig:5}b). This feature is sensitive to the energetic-scale motions as the entropy of an IAAFT surrogate signal (pink dash-dotted line) shows a similar inflection as the original one. Therefore, to investigate this phenomenon more carefully, we evaluate the normalized Shannon entropies of $N_{p}$ separately for the positive ($\mathcal{H}^{u^{\prime}>0}_{n}(N_{p})$) and negative ($\mathcal{H}^{u^{\prime}<0}_{n}(N_{p})$) fluctuations. The computation of $\mathcal{H}^{u^{\prime}>0}_{n}(N_{p})$ or $\mathcal{H}^{u^{\prime}<0}_{n}(N_{p})$ is similar to Eq. \ref{SN}, where the event lengths and their probabilities are conditioned on positive or negative fluctuations. Unlike SLTEST, for the TBL experiment, $P(t_{p}^{+})$ displays a distinctly different behavior between $u^{\prime}>0$ and $u^{\prime}<0$ signals. For instance, heavy tails of the event time scale PDFs (quantified through the kurtosis of $t_{p}$) are governed by the negative events as compared to the positive ones (see Fig. S2 in the supplementary material). Note that this difference is not reflected in the mean time scale ($\overline{{t_{p}}^{+}}$) and is only evident through the large-scale events (Fig. S2a--b).

Coming back to Fig. \ref{fig:5}b, we observe the inflection in $\mathcal{H}^{u^{\prime}}_{n}(N_{p})$ is captured in the negative events (light-blue line) as opposed to the positive ones (cyan line). More importantly, beyond $z^{+}=70$, $\mathcal{H}^{u^{\prime}>0}_{n}(N_{p})$ approaches a near-constant value. This indicates the organizational structure of the high-speed streaks ($u^{\prime}>0$) is height-invariant at $z^{+}>70$. Recent numerical experiment results of \citet{bae2021life} show that the low-speed ($u^{\prime}<0$) streaks in wall-bounded flows merge progressively as the heights increase from the viscous sublayer to the inertial layer. They conclude that the low-speed streaks change their characteristics at approximately $z^{+}=70$, the same location where we observe the inflection point in $\mathcal{H}^{u^{\prime}}_{n}(N_{p})$. Therefore, this inflection can be interpreted as a sign of the change in the structural properties of turbulence as one transitions from the viscous sublayer to the inertial or log layer. We next demonstrate how these coherent structures influence the temporal evolution of the signal by investigating the relationship between $S^{2}_{p}$ and $t_{p}$.

\subsubsection{Event contributions}
In Fig. \ref{fig:5}c, we show the contour plot of normalized event contributions to the streamwise velocity variance ($S^{2+}_{p,u}$) against the event time scales $(t_{p,u}\overline{u})/\delta$ and heights ($z^{+}$) from the TBL experiment. Note that we convert $t_{p}$ to a length scale using the local mean wind speed ($\overline{u}$), and subsequently normalize it with $\delta$. Through such scaling, we intend to probe the influence of outer-scale structures on event statistics. 

The event contributions are converted to densities by dividing them with the logarithmic bin-width of $(t_{p,u}\overline{u})/\delta$ so that when integrated over all the $(t_{p,u}\overline{u})/\delta$ values the result is ${\sigma_{u}}^2/u_{*}^2$. We denote these event densities as $\langle S^{2+}_{p,u} \vert \frac{(t_{p,u}\overline{u})}{\delta} \rangle$ and their logarithms are plotted as the contours in Fig. \ref{fig:5}c. The individual event contribution curves at each $z^{+}$ value are shown in Fig. S3b of the supplementary material, whose maxima are highlighted through green triangle markers. The black contour lines in Fig. \ref{fig:5}c denote the regions of substantial contributions to $\sigma_{u}^2$ from some specific events. On the individual event contribution curves (Fig. S3b), these specific events are demarcated by two black dash-dotted horizontal lines. The blue vertical lines in Fig. \ref{fig:5}c indicate the locations of inner- and outer-spectral peak positions in outer-layer coordinates \citep{baars2015wavelet}. On the other hand, the two horizontal lines in Fig. \ref{fig:5}c represent those $z^{+}$ locations where the inner- ($z^{+}=12$) and outer-spectral ($z^{+}=474$) peak positions appear \citep{baars2015wavelet}. 

If one locates those $(t_{p,u}\overline{u})/\delta$ values corresponding to which the event contributions are maximum ($[(t_{p}\overline{u})/\delta]_{\rm max}$) and plot their vertical profiles (shown as green triangles in Fig. \ref{fig:5}c), they follow a distinct power-law of $+1.6$. This is apparently more clear in Fig. \ref{fig:5}d, where a $+1.6$ power-law is fitted to the green triangles. Furthermore, $[(t_{p}\overline{u})/\delta]_{\rm max}$ approach the outer-spectral peak position as the height increases and happen to be nearly equal to the integral scale of $u^{\prime}$ ($\gamma_{u}$, black dash-dotted line in Fig. \ref{fig:5}d). As a standard practice, $\gamma_{u}$ is obtained by integrating the auto-correlation function up to its first zero-crossing \citep{li2012mean}. 

Similar to $[(t_{p}\overline{u})/\delta]_{\rm max}$, the black contour lines of $S^{2+}_{p,u}$ vertically evolve in a power-law fashion, i.e., they vary as ${(z^{+})}^{1.6}$. Note that the power-law portion of the black contour lines is only evident beyond a certain $z^{+}$, approximately where the logarithm region starts. This power law is shown as a pink line with circular markers in Fig. \ref{fig:5}c. From Fig. \ref{fig:5}d, one also notices that the maximum event contributions (red line with triangles) match with the vertical profile of ${\sigma_{u}}^2/u_{*}^2$ (light blue line). 

Particularly for the logarithmic layer, the vertical profile of ${\sigma_{u}}^2/u_{*}^2$ is predicted by the attached eddy hypothesis \citep{marusic2019attached}, and therefore, these results imply that most of the event contributions come from such coherent structures. However, in an event-based framework, self-similarity of the attached eddies in the vertical direction is imposed as a ${(z^{+})}^{1.6}$ power-law instead of just $z^{+}$. The expectation of $z^{+}$ scaling arises from how the frequency spectra of streamwise velocity signals scale with height in the logarithmic region of wall-bounded flows \citep{jimenez2012cascades,baars2020data}. In the spectral representation, the attached eddies are assumed to be space-filling \citep{marusic2019attached}. Yet, from Fig. \ref{fig:5}c, it is evident that the black contour lines deviate significantly from a $+1$ power law as indicated by the grey dashed line. We hypothesize that this distinction arises because in an event framework, the attached eddies need not be space-filling, and accordingly, they can be a part of a fractal set with a non-integer dimension. This is at present a conjecture, and further pursuance of it is beyond the scope of this study.  

In addition to the TBL experiment, one observes an almost identical behavior if the SLTEST dataset is considered. For instance, in Fig. \ref{fig:5}e, the vertical evolution of the normalized event contributions towards $\sigma_{u}^2$ and $\sigma_{w}^2$ are shown. These event contributions are represented through filled contours for $\sigma_{u}^2$ while the contour lines represent $\sigma_{w}^2$. Although $\delta$ was not directly available at the SLTEST site, we used the integral scale of $u^{\prime}$ at the topmost SLTEST height as its proxy. In Fig. \ref{fig:5}f, the contours are shown for the momentum flux. Here we consider the absolute momentum flux signal $\vert u^{\prime}w^{\prime} \vert$ while describing the event features. From both Fig. \ref{fig:5}e--f it is clear that the significant event contributions do vertically evolve as a ${(z^{+})}^{1.6}$ power-law (shown as a pink line with circular markers). Since large event contributions are associated with strong amplitude variations it is interesting to see how such behavior is encoded in the burstiness index. 
 
\subsubsection{Burstiness behavior}
In Fig. \ref{fig:5}g, we show the vertical profiles of burstiness indices corresponding to the instantaneous evolution of velocity variances ($B^{2}_{u}$ and $B^{2}_{w}$) and absolute momentum flux ($B^{1}_{u^{\prime}w^{\prime}}$) signals. It is clear that the behavioral features of this index are nearly indistinguishable among all the flow quantities with all showing an increase with height. This outcome is very different from the perspective of signal PDFs as those are considerably different for the three flow quantities (see Fig. \ref{fig:10} in Appendix \ref{app_C}). Furthermore, in contrast to the signal PDFs, the burstiness index changes for an RS time series (light-pink line) but remains nearly preserved in an IAAFT surrogate (dark-pink line). This is demonstrated through the $u^{\prime}$ signal from the TBL experiment. For this signal, the vertical profile of $B^{2}_{u}$ of an RS sequence is qualitatively similar to excess kurtosis ($\mathcal{K}_e$) in Fig. \ref{fig:10} (Appendix \ref{app_C}). However, as soon as the energy spectrum of the signal is considered through an IAAFT surrogate, the burstiness index becomes almost equal to the original one. Therefore, this index, unlike kurtosis, takes the coherent structures into account while quantifying strong amplitude variations in the signal.

Hitherto, we have focused on the full-signal behavior while discussing the bursts in the generation of velocity variance or momentum flux. As discussed above, these bursts are typically related to the presence of attached eddies in the logarithmic layer. However, it is not immediately clear how these bursts are different from the small-scale bursts which cause large amplitude fluctuations in velocity increments (e.g., $\Delta u(\tau)$). Accordingly, one may ask if the small-scale bursts are more intense than the ones associated with ${x^{\prime}}^2$ ($x=u,w$) or $u^{\prime}w^{\prime}$ signals. To investigate such aspects, we introduce a scale-dependent event framework. The associated technical details are illustrated through an example in Sect. \ref{method}. Below we describe the results obtained from this framework.     

\subsection{A scale-dependent event framework}
\label{scale_event}
\begin{figure}[h]
\centering
\hspace*{-0.6in}
\includegraphics[width=1.25\textwidth]{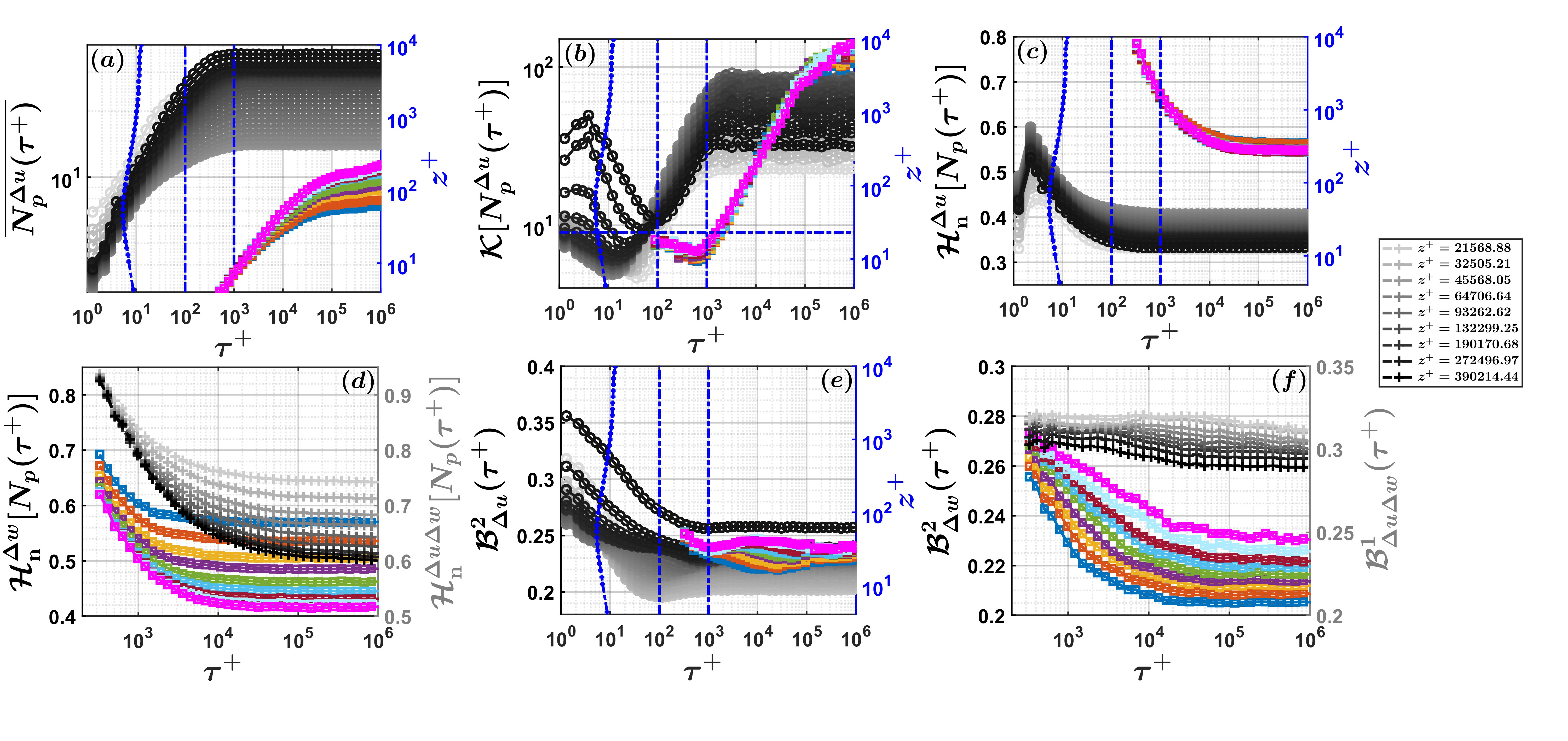}
  \caption{The scale-dependent (a) mean event lengths ($\overline{N^{\Delta u}_{p}(\tau^{+})}$), (b) kurtosis of event lengths ($\mathcal{K}[N^{\Delta u}_p({\tau}^{+})]$), (c) normalized Shannon entropy of event lengths ($\mathcal{H}^{\Delta u}_{n}[N_{p}(\tau^{+})]$) are plotted against $\tau^{+}$ for the horizontal velocity increments computed from the TBL and SLTEST datasets. The color codes are similar to Fig. \ref{fig:4}d--f. In (a)--(c) and (e), the right-hand-side of the $y$-axis represents the vertical profile of the inner-scaled Taylor-microscale ($\lambda^{+}$) evaluated from the $u^{\prime}$ signal of the TBL dataset (blue-dotted line). In (d), the normalized Shannon entropy of event lengths is shown for the vertical velocity ($\Delta w$) and mixed-order increments ($\Delta u \Delta w$) from the SLTEST dataset. The mixed-order increments are represented at the right-hand side of $y$-axis and grey color shades are used to denote the nine SLTEST heights (see the legend at the extreme right end). (e) The scale-dependent burstiness indices are shown for the horizontal velocity increments. (f) Burstiness indices corresponding to $\Delta w$ and $\Delta u \Delta w$ are shown from the SLTEST dataset. Similar to (d), the mixed-order increments are represented by the right-hand side of the $y$-axis, albeit the original values are vertically shifted for clarity purposes.}
\label{fig:6}
\end{figure}
Through this scale-dependent event framework, we first demonstrate a statistical correspondence between the eddy and event time scales ($t_{p}$). This is often considered to be a challenging issue since in event analysis the structures are based in physical space while the eddy time scales are generally represented through Fourier modes \citep{sreenivasan1991local}. However, through the Wiener-Khinchin theorem, since the structure functions are equivalent to the Fourier spectra, the eddy time scales can also be defined in terms of time lags or $\tau$. For our purposes, we normalize $\tau$ with wall-unit scaling and denote it as $\tau^{+}$. To highlight any height-dependence, we prefer to use $\tau^{+}$ instead of converting the same to the spatial lags. Subsequently, for each $\tau^{+}$, one computes the event statistics of the velocity increments (e.g., $\Delta u (\tau^{+})$). For instance, similar to Fig. \ref{fig:1}a, one can define $N_{p}$ (event lengths) values for the $\Delta u (\tau^{+})$ signal. If with increasing $\tau^{+}$, the event statistics converge towards the values as obtained from the full signal (e.g., $u^{\prime}$, $w^{\prime}$, or $u^{\prime}w^{\prime}$), one can infer the PDFs of event time scales ($P({t_{p}}^{+})$, Fig. \ref{fig:5}a) is a cumulative effect of all the eddy structures present in the flow. By doing so, one establishes an association between the eddy and event time scales.

\subsubsection{Correspondence between eddy and event time scales}
To accomplish this objective, we choose the mean and kurtosis of event lengths ($N_{p}$) as the two relevant statistical measures. Physically, mean event length ($\overline{N_{p}}$) is inverse of the zero-crossing density, a quantity which is often linked to the Taylor microscale \citep{sreenivasan1983zero,poggi2009flume}. On the other hand, kurtosis of event lengths ($\mathcal{K}(N_{p})$) is related to how fat the tails of the event PDFs are. 

In Figs. \ref{fig:6}a--b, we present $\overline{N_p({\tau}^{+})}$ and $\mathcal{K}[N_p({\tau}^{+})]$ for the $\Delta u ({\tau}^{+})$ signals from the TBL and SLTEST experiments. For comparison purposes, we mark the kurtosis of an exponential distribution ($\mathcal{K}=9$) in Fig. \ref{fig:6}b, i.e., the distribution of disordered event lengths. One can clearly see, $\overline{N_p^{\Delta u}({\tau}^{+})}$ and $\mathcal{K}[N_p^{\Delta u}({\tau}^{+})]$ indeed attain a plateau towards the full signal values (evident from the flat regions) as the large-scale structures are considered. To be precise, for the TBL dataset, this saturation occurs at time scales nearly equal to the outer spectral peak position, which is at $\tau^{+}=1000$. Therefore, one can conclusively prove that the heavy tails of the event time scale PDFs in Fig. \ref{fig:5}a are a result of the large-scale structures (comparable to the outer-layer scales) passing over the measurement location. 

Interestingly, $\overline{N_p^{\Delta u}({\tau}^{+})}$ values increase monotonically with $\tau^{+}$ whereas for $\mathcal{K}[N_p^{\Delta u}({\tau}^{+})]$ a monotonic increase is observed only beyond the inner spectral peak position, i.e, for $\tau^{+}>100$. In fact, barring the top four heights of the TBL dataset ($z^{+}=$ 5610--9965), $\mathcal{K}[N_p^{\Delta u}({\tau}^{+})]$ undergoes a transformation from sub-exponential ($\mathcal{K}<9$) to super-exponential ($\mathcal{K}>9$) distribution as one crosses $\tau^{+}=100$. Apart from this, the result related to $\overline{N_p^{\Delta u}({\tau}^{+})}$ presents a contradiction with previous studies. For instance, \citet{sreenivasan1983zero} interpreted the mean zero-crossing density of a turbulent signal to be proportional to the Taylor microscale by using a formulation proposed by \citet{rice1945mathematical}. For a turbulent time series (lets say $u^{\prime}$), the Taylor microscale ($\lambda$) is defined as,
\begin{equation}
    \lambda=\frac{\sigma_{u}}{{\Bigg[\overline{{\Big(\frac{\partial u}{\partial t}\Big)^2}}\Bigg]}^{\frac{1}{2}}},
\end{equation}
which physically represents the time scales of the dissipative structures \citep{katul1995local}. Since mean event length is an inverse of zero crossing density, one would thus expect $\overline{N_p^{\Delta u}({\tau}^{+})}$ will converge at scales comparable to $\lambda^{+}$ (scaled with wall-units). However, such expectation does not hold, as one could see from Fig. \ref{fig:6}a that $\overline{N_p^{\Delta u}({\tau}^{+})}$ converge at scales $\tau^{+}=1000$, which is many orders larger than $\lambda^{+}$ (shown as a blue line from the TBL dataset). As a consequence, this negates any possibility of associating the mean zero-crossing density to $\lambda$. Note that we only compute $\lambda^{+}$ from the TBL dataset given its fine temporal resolution of the order of Kolmogorov scales. 

Moreover, this framework can even be extended to study the organizational features of turbulence at each scale of the flow. In fact, similar to using the normalized Shannon entropy (with respect to an RS signal) of $N_p$ ($\mathcal{H}_{n}(N_{p})$), one can also investigate the scale-wise evolution of this quantity by extending it to the velocity increments. Namely, one can use the same Eq. \ref{SN} to compute $\mathcal{H}_{n}(N_{p})$ but for the $\Delta u$ signal at any time-lag $\tau^{+}$. In Fig. \ref{fig:6}c, we plot $\mathcal{H}_{n}[N_{p}(\tau^{+})]$ of $\Delta u$ signal from the TBL and SLTEST experiments. For the TBL dataset, one can notice that, irrespective of $z^{+}$, the maximum values of $\mathcal{H}^{\Delta u}_{n}[N_{p}(\tau^{+})]$ appear at around $\tau^{+} \approx 5$. Since this peak time scale of the Shannon entropy is comparable to $\lambda^{+}$, the dissipative structures (identified through $\lambda^{+}$) are more disorganized as compared to the rest of the scales.
Nevertheless, as the scales increase $\mathcal{H}^{\Delta u}_{n}[N_{p}(\tau^{+})]$ decreases (thereby indicating more organization) and eventually saturates towards $\mathcal{H}^{u^{\prime}}_{n}(N_{p})$. 

A similar situation is observed with the SLTEST dataset, i.e., the values of $\mathcal{H}^{\Delta u}_{n}[N_{p}(\tau^{+})]$ decrease with the increasing time scales. However, the $\mathcal{H}^{\Delta u}_{n}[N_{p}(\tau^{+})]$ values of the SLTEST dataset remain substantially larger than the TBL one. Therefore, this implies a $Re$-dependence on how the eddy structures organize themselves at each $\tau^{+}$. Although at inertial subrange scales the turbulence features are assumed to be $Re$-independent (owing to local isotropy assumption), our results indicate that this does not hold for the present datasets at hand. Additionally, $\mathcal{H}^{\Delta u}_{n}[N_{p}(\tau^{+})]$ curves display an excellent collapse for all the nine heights of the SLTEST data. Previous studies have indicated that the outer-layer structures (scale with the boundary-layer depth, $\delta$) govern the organizational features of streamwise velocity fluctuations in atmospheric surface layer flows \citep{drobinski2004structure}. One plausible interpretation of this collapse is these global structures not only determine the large-scale organizational features of $u^{\prime}$ but extend their footprints down to inertial subrange eddies. Evidently, presence of such large-scale structures violates the principle of local isotropy, which by the way is also reflected in the $D_{ww}/D_{uu}$ ratios being smaller than 4/3. 

On the contrary to $\mathcal{H}^{\Delta u}_{n}[N_{p}(\tau^{+})]$, a different scenario arises for $\Delta w$ and $\Delta u \Delta w$ signals. Similar to the vertical profiles of $\mathcal{H}^{w^{\prime}}_{n}(N_{p})$ and $\mathcal{H}^{u^{\prime}w^{\prime}}_{n}(N_{p})$ (Fig. \ref{fig:5}b), $\mathcal{H}^{\Delta w}_{n}[N_{p}(\tau^{+})]$ and $\mathcal{H}^{\Delta u \Delta w}_{n}[N_{p}(\tau^{+})]$ show a clear height dependence across all $\tau^{+}$ values (Fig. \ref{fig:6}d). For visualization purposes, $\mathcal{H}^{\Delta u \Delta w}_{n}[N_{p}(\tau^{+})]$ of the SLTEST dataset are shown on the right-hand-side axis of Fig. \ref{fig:6}d with the heights being identified in grey-shaded colors (see the legend). Given the local attached eddies (scales with $z$) have height-dependent features, this result indicates that they exert their influence at scales comparable to the inertial subrange scales. However, at inertial subrange scales, since negligible transport of momentum is accomplished (Fig. \ref{fig:4}f and \ref{fig:9}d), they mostly act as inactive motions \citep{bradshaw1967inactive}.

\subsubsection{Scale-dependent burstiness index}
The results presented so far illustrate a close association between the eddy and event time scales. After establishing such a connection, we next evaluate the scale-wise evolution of the burstiness index. We focus on the second- and mixed-order velocity increments since these two quantities describe the scale-wise contributions to velocity variances ($\sigma_{x}^2$, $x=u,w$) and momentum fluxes ($\overline{u^{\prime}w^{\prime}}$). In Fig. \ref{fig:6}e we show $\mathcal{B}^2_{\Delta u}(\tau^{+})$ from the TBL and SLTEST experiment. For the TBL data, $\mathcal{B}^2_{\Delta u}(\tau^{+})$ decreases as the scales increase with the largest values being typically associated with the dissipative structures. Eventually, at scales $\tau^{+}=1000$ and beyond, $\mathcal{B}^2_{\Delta u}(\tau^{+})$ approach towards the full signal value, which is $\mathcal{B}^2_{u}$ (shown in Fig. \ref{fig:5}g). 

Furthermore, much like the vertical profile of $\mathcal{B}^2_{u}$, the shapes of $\mathcal{B}^2_{\Delta u}(\tau^{+})$ curves change with height, thereby implying a connection between the small- and large-scale bursts. Typically, the influence of large scales on small-scale statistics is hypothesized to be the reason behind the appearance of anomalous scalings in structure function moments \citep{ecke2005turbulence}. It is therefore encouraging to notice that the scale-dependent burstiness index captures such information quite seamlessly by only considering the ${\Delta u}^2$ signal. However, a weak scale dependency in $\mathcal{B}^2_{\Delta u}(\tau^{+})$ is noted for the SLTEST data. A similar inference is obtained if one investigates the mixed-order velocity increments. 

For instance, in Fig. \ref{fig:6}f, on the right-hand-side axes, $\mathcal{B}^1_{\Delta u \Delta w}(\tau^{+})$ are plotted from the SLTEST data. To better clarify the features, curves are slightly vertically shifted and the heights are grey-shaded (see the legend). Identical to $\mathcal{B}^2_{\Delta u}(\tau^{+})$, no significant scale-wise variations are noted in this quantity for any measurement level. This occurs in spite of the presence of a clear inertial subrange in the second- ($\overline{{\Delta u}^2}$) and mixed-order ($\overline{\Delta u \Delta w}$) structure functions (see Fig. \ref{fig:4}d and f). Therefore, for the atmospheric flows, the role of strong amplitude fluctuations or bursts in the generation of streamwise velocity variances or momentum fluxes remains nearly equal across all the eddy time scales. Moreover, an identical situation prevails if one considers the cross-stream components, such as $\mathcal{B}^2_{\Delta v}(\tau^{+})$ and $\mathcal{B}^1_{\Delta v \Delta w}(\tau^{+})$ (see Fig. S4 in supplementary material). Nevertheless, the same is not true concerning $\mathcal{B}^2_{\Delta w}(\tau^{+})$ (Fig. \ref{fig:6}f, left-hand-side axes). 

In fact, $\mathcal{B}^2_{\Delta w}(\tau^{+})$ values not only display a scale dependence but also vary with height. More importantly, although the scale-wise organizational features of $\Delta w$ and $\Delta u \Delta w$ signals remain qualitatively similar (see Fig. \ref{fig:6}d), their burst characteristics ($\mathcal{B}^2_{\Delta w}(\tau^{+})$ and $\mathcal{B}^1_{\Delta u \Delta w}(\tau^{+})$) are significantly different. Instead of following $\mathcal{B}^2_{\Delta w}(\tau^{+})$, the scale-wise variations in $\mathcal{B}^1_{\Delta u \Delta w}(\tau^{+})$ follow the same trend as in $\mathcal{B}^2_{\Delta u}(\tau^{+})$. Unfortunately, due to the unavailability of $w^{\prime}$ data, the conclusions regarding the vertical velocity and mixed-order increments cannot be validated for the TBL dataset.

\begin{figure}[h]
\centering
\hspace*{-1.2in}
\includegraphics[width=1.6\textwidth]{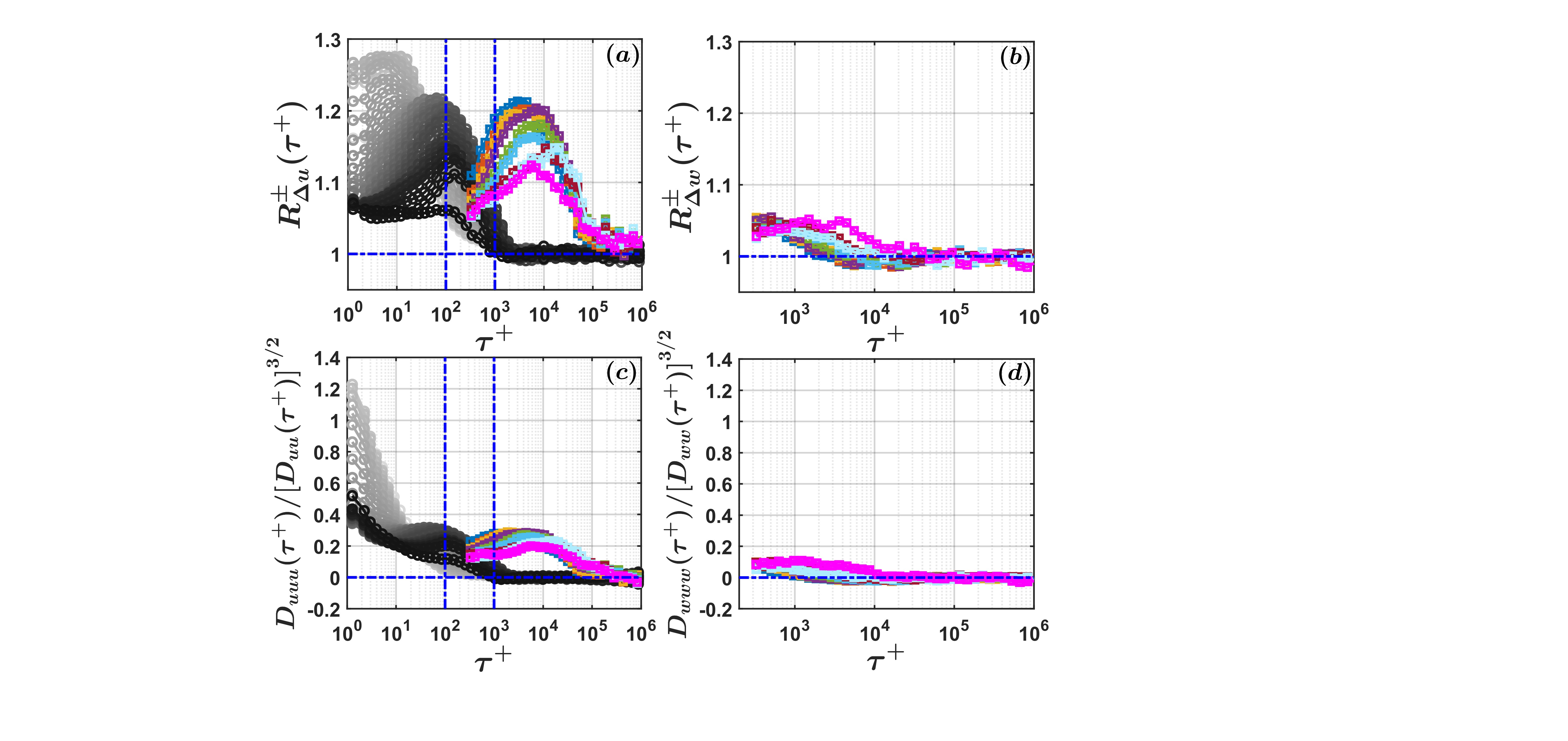}
  \caption{The ratios between the burstiness indices computed for positive and negative values (see Eq. \ref{BI_ratio}) of (a) $\Delta u$ ($R^{\pm}_{\Delta u}(\tau^{+})$) and (b) $\Delta w$ ($R^{\pm}_{\Delta w}(\tau^{+})$), are plotted against $\tau^{+}$. The horizontal blue line indicates unity, i.e. when the positive and negative velocity increments have similar burstiness features. In (c)--(d), the skewness of the velocity structure functions ($D_{xxx}/{(D_{xx})}^{3/2}$, $x=u,w$) are shown corresponding to the $u^{\prime}$ and $w^{\prime}$ signals, respectively. The zero skewness is denoted by the blue horizontal lines and the color codes are the same as in Fig. \ref{fig:4}d--f.}
\label{fig:7}
\end{figure}

At a first glance, these burst results seem to paint a counter-intuitive picture. One would expect the burstiness activities to increase as the scales decrease due to the presence of large non-Gaussian fluctuations which cause small-scale intermittency \citep{sreenivasan1997phenomenology}. One of the aspects of non-Gaussianity is a statistical asymmetry between the positive and negative values \citep{singh2016measure}. Whether the scale-dependent event framework captures such non-Gaussian aspects, one can investigate the burstiness index separately for the positive and negative velocity increments. For carrying out this computation, one first conditions the event lengths and sizes based on the sign of the velocity increments. Thereafter, the burstiness curves are plotted separately for the positive and negative increments with the indices (e.g., $\mathcal{B}^2_{\Delta u>0}(\tau^{+})$, $\mathcal{B}^2_{\Delta u<0}(\tau^{+})$) being calculated as per the procedure described in Fig. \ref{fig:2}a. 

To quantify any asymmetry, a ratio between the two is obtained and denoted as,
\begin{equation}
R^{\pm}_{\Delta x}(\tau^{+})=\frac{\mathcal{B}^2_{\Delta x>0}(\tau^{+})}{\mathcal{B}^2_{\Delta x<0}(\tau^{+})}, \ x=u,w.
\label{BI_ratio}
\end{equation}
In Fig. \ref{fig:7} we present these ratios and structure-function skewness of $u^{\prime}$ and $w^{\prime}$ signals from both experiments. The non-zero values of the structure-function skewness, $D_{xxx}(\tau^{+})/[D_{xx}(\tau^{+})]^{3/2}$ with $x=u,w$, is a measure of non-Gaussianity of small-scale turbulence, where the notation $D_{xxx}(\tau^{+})$ denotes the third-order structure function, i.e., $\overline{[\Delta x(\tau^{+})]^3}$. On the other hand, if $R^{\pm}_{\Delta x}(\tau^{+})$ are equal to unity, no asymmetry exists between the burstiness features of positive and negative velocity increments. One could observe, regarding $\Delta u(\tau^{+})$, $R^{\pm}_{\Delta u}(\tau^{+})$ and $D_{uuu}(\tau^{+})/[D_{uu}(\tau^{+})]^{3/2}$ behave similarly, with both showing a significant deviation from unity or zero (depending on the statistic) as the scales decrease (Fig. \ref{fig:7}a and c). Moreover, as opposed to $\mathcal{B}^2_{\Delta u}(\tau^{+})$, the variations in $R^{\pm}_{\Delta u}(\tau^{+})$ remain remarkably identical between the SLTEST and TBL datasets. 

In fact, for both of these datasets, $R^{\pm}_{\Delta u}(\tau^{+})$ attains a clear peak at some intermediate scales. Specific to the TBL dataset, this peak corresponds to the inner-spectral peak position ($\tau^{+}=100$) for the heights within the logarithmic layer. However, as one approaches the viscous sublayer, large values of $R^{\pm}_{\Delta u}(\tau^{+})$ are typically associated with scales comparable to $\lambda^{+}$. Eventually, at larger scales ($\tau^{+}>1000$ for TBL dataset), both $R^{\pm}_{\Delta u}(\tau^{+})$ and $D_{uuu}(\tau^{+})/[D_{uu}(\tau^{+})]^{3/2}$ saturate to unity and zero, respectively. Therefore, $R^{\pm}_{\Delta u}(\tau^{+})$ successfully captures the non-Gaussian features of small-scale turbulence. Additionally, the asymmetry between the positive and negative velocity increments at smaller scales of the flow is also reflected in their organizational structure as confirmed by the entropy ratios $\mathcal{H}^{\Delta u>0}_{n}/\mathcal{H}^{\Delta u<0}_{n}$ being greater than 1 (see Fig. S5a in the supplementary material). In contrast, for $\Delta w$ signal, no such asymmetry is noted in $R^{\pm}_{\Delta w}(\tau^{+})$, structure function skewness, or in their entropy ratio $\mathcal{H}^{\Delta w>0}_{n}/\mathcal{H}^{\Delta w<0}_{n}$  (Fig. \ref{fig:7}b and d, Fig. S5b). The vanishing skewness of $\Delta w$ signal appears to be in agreement with the results of \citet{mestayer1982local} from a high-$Re$ boundary layer flow. 

In summary, the scale-dependent event framework provides very useful information about the structural properties of turbulence at both small and large scales of the flow. Notwithstanding the non-Gaussian features of small-scale turbulence (in terms of skewness) is identified through this framework, an interesting result emerges when one considers the scale-wise evolution of burstiness indices related to $\Delta u$ and $\Delta u \Delta w$ signals. As opposed to the TBL dataset, the variations in $\mathcal{B}^2_{\Delta u}(\tau^{+})$ and $\mathcal{B}^1_{\Delta u \Delta w}(\tau^{+})$ of the atmospheric flow are found to be nearly scale-invariant. Physically this finding implies, at smaller scales of a near-neutral atmospheric flow, the connection between burst-like activities and small-scale intermittency is not straightforward. On a more fundamental level, the $Re$-dependence in the behavior of the burstiness index at smaller scales of the flow bears a resemblance with the results of \citet{yeung2015extreme}. Through direct numerical simulations, \citet{yeung2015extreme} pointed out that the features of large-amplitude events of small-scale turbulence do not necessarily scale with the Reynolds number of the flow. It is promising to note that our results confirm their prediction, although through a time-series analysis with limited spatial information in the vertical direction. A consequence of such limitation is, it is at present unclear how exactly the three-dimensional flow structures induce a $Re$-dependence on the small-scale turbulent bursts, therefore requiring further research. We present our conclusions in the next section.

\section{Conclusion}
\label{conclusion}
In this study, we propose a novel scale-dependent event framework that enables us to quantify the role of strong amplitude fluctuations (or bursts) in turbulence generation across multiple eddy time scales. To be specific, we intend to probe whether the generation of turbulence at smaller scales of the flow appears more bursty than at larger scales. To achieve this objective, we revisit the "burstiness index" and apply it to the velocity fluctuations and their increments. Our approach is in contrast with previous research where the event framework had mainly been employed to investigate the strong events in velocity fluctuations rather than their increments. In particular, through our approach, we establish a linkage between the small- and large-scale bursts in wall-bounded turbulent flows. Moreover, we compare our findings between two experiments conducted in a wind tunnel and in a near-neutral atmosphere (without buoyancy) with the Reynolds number ($Re$) being different by almost two orders of magnitude.

Through this framework, we first demonstrate how the organizational structures of the two flows vary by exploiting a new metric based on the Shannon entropy of event lengths. We find that in both flows, notwithstanding their different organization, burst-like features in the instantaneous velocity variances (${u^{\prime}}^2(t)$, ${w^{\prime}}^2(t)$) and momentum flux ($u^{\prime}w^{\prime}(t)$) signals are governed by the coherent structures. Particularly, for heights within the logarithmic layer, these coherent structures are best represented by the attached eddies. However, unlike the spectral prediction, our evidence suggests that the attached eddies in an event framework are identified through a non-integer power-law of height, i.e., $z^{1.6}$. Besides, when the burst characteristics of ${u^{\prime}}^2(t)$, ${w^{\prime}}^2(t)$, and $u^{\prime}w^{\prime}(t)$ signals are compared with each other, they are found to be remarkably similar. On the other hand, a dissimilarity among these three variables is observed when one considers the scale-wise evolution of their burstiness indices. Therefore, to further illustrate how these bursts associated with coherent structures are different from the bursts at smaller scales of the flow (inertial subrange and dissipative range), a statistical correspondence is established between the eddy and event time scales. While doing so, an intriguing scenario appears by turning one's attention towards small-scale bursts.

Despite the non-Gaussian aspects (considering only skewness) of small-scale turbulence captured through the scale-dependent event framework, a $Re$-dependence is noted while studying how the burstiness characteristics of the Reynolds stress components evolve across different scales of the flow. In this context, the scale-wise generation of the Reynolds stress components are described through second-order (${\Delta u}^2$, ${\Delta w}^2$) and mixed-order ($\Delta u \Delta w$) velocity increments, respectively. Regarding the wind-tunnel dataset at an $Re \approx 14750$, we find that the generation of streamwise velocity variances become progressively more bursty as the eddy time scales decrease. On the other hand, for atmospheric flows at an ultra-high Reynolds number ($Re \approx 10^{6}$), the burstiness features of ${\Delta u}^2$ and $\Delta u \Delta w$ signals are found to be approximately scale-invariant. In contrast, ${\Delta w}^2$ signals display strong burst-like features as the eddy time scales decrease. Thus, for high-$Re$ flows, as opposed to general perception, a non-trivial relationship exists between small-scale intermittency and burst-like activities in the turbulent signal.

Undoubtedly, these results open up new research avenues. For instance, one could ask, why in the case of atmospheric flows the burst features of streamwise velocity variances and momentum fluxes remain nearly equal across all the eddy time scales? How such a phenomenon connects with small-scale intermittency and what is the effect of buoyancy on this? Would the effect of bursts be similar if different scalar fluctuations and their fluxes are considered? What is the role of the underlying surface, such as a canopy, on burstiness? We leave these questions for our future research. 

\section*{Acknowledgements}
We dedicate this study to the memory of late Prof. Roddam Narasimha. SC acknowledges the Department of Civil and Environmental Engineering, UC Irvine, for providing the financial support. The authors would like to thank Dr. KG McNaughton for providing them the SLTEST dataset. SC thanks Dr. Giovanni Iacobello for some initial help with the processing of wind-tunnel dataset. The wind-tunnel experiment data are available at \url{https://doi.org/10.26188/5e919e62e0dac}. TB acknowledges the funding support from the University of California Office of the President (UCOP) grant LFR-20-653572 (UC Lab-Fees); the National Science Foundation (NSF) grants NSF-AGS-PDM-2146520 (CAREER), NSF-OISE-2114740 (AccelNet) and NSF-EAR-2052581 (RAPID); the United States Department of Agriculture (USDA) grant 2021-67022-35908 (NIFA); and a cost reimbursable agreement with the USDA Forest Service 20-CR-11242306-072.  

\appendix
\appendixpage
\begin{appendices}
\section{Random-shuffling and phase-alteration experiments}
\label{app_A}

\begin{figure}[h]
\centering
\hspace*{-0.6in}
\includegraphics[width=1.6\textwidth]{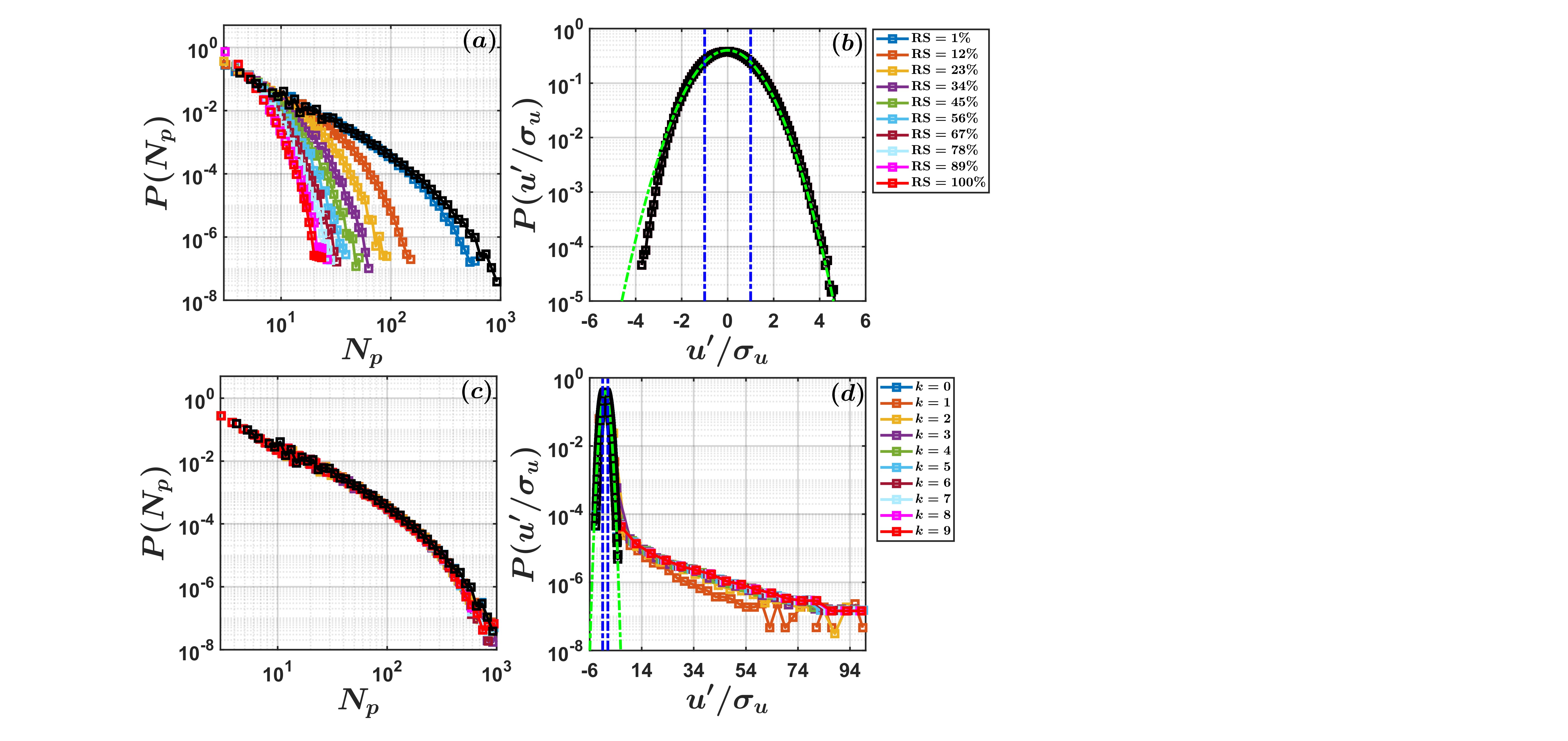}
  \caption{(a) The PDFs of event lengths ($N_{p}$) are shown for the $u^{\prime}$ signal at $z^{+}=67$, by gradually increasing the strength of the random-shuffling (RS) from 1 to 100 $\%$. The colored lines correspond to different RS strengths (see the legend), while the black line represents the original $u^{\prime}$ signal. (b) The PDFs of normalized velocity fluctuations ($u^{\prime}/\sigma_{u}$) are shown for the original and randomly-shuffled $u^{\prime}$ signals. Two vertical dash-dotted blue lines represent $u^{\prime}/\sigma_{u}=\pm 1$, and the green dash-dotted line indicates the Gaussian distribution. Similar to (a), in (c) we show the PDFs of $N_{p}$ by gradually altering the Fourier phase angle distributions of $u^{\prime}$ through sampling them from a von Mises distribution with a parameter $k$. The colored lines correspond to different $k$ parameters (see the legend). (d) The PDFs of $u^{\prime}/\sigma_{u}$ are shown for the $u^{\prime}$ signals with different $k$ parameters and the original one.}
\label{fig:8}
\end{figure}

We explain the methodologies to create two different surrogate signals, one of which preserves the signal PDFs but alters the PDFs of event lengths ($N_{p}$) whereas for the other, the PDFs of $N_{p}$'s are preserved but the signal PDFs are changed. The first of such surrogate signals are generated through gradual random-shuffling, while for the latter a phase-alteration technique is used. 

In a gradual random-shuffling method, we first choose any signal, for instance the $u^{\prime}$ signal at $z^{+}=$ 67, and then locate the midpoint of the signal which will be at $N/2$'th point if the signal length is $N$. Thereafter, to create a randomized dataset at an $x$\% randomization strength (RS), $x/2$\% of the time series values are randomly shuffled between the left and right halves, i.e., along the midpoint of the time series. By doing so, we progressively destroy the temporal coherence in the signal (thereby altering the event lengths) but preserve the signal PDF since the time series values remain the same. In Fig. \ref{fig:8}a--b, we illustrate this by showing the PDFs of $N_{p}$ and $u^{\prime}/\sigma_{u}$. One can clearly notice, $P(N_{p})$ varies greatly for different values of RS while $P(u^{\prime}/\sigma_{u})$ is unchanged. 

To generate the second type of surrogates, Fourier phase distributions of a signal are altered through a phase alteration experiment. To achieve this objective, one first takes the Fourier transform of a signal and then computes the amplitudes and phases of the Fourier coefficients. As a next step, the Fourier amplitudes are kept the same but its phases are sampled from a different distribution than the original one. After altering the phases, one eventually takes an inverse Fourier transform to generate a surrogate dataset. By preserving the Fourier amplitudes, surrogate datasets from phase alteration experiment share the same Fourier spectrum or the auto-correlation function as the original. This ensures the PDFs of event lengths remain identical since those are sensitive to the auto-correlation function of the time series \citep{majumdar1999persistence,chamecki2013persistence}. On the other hand, the alteration of Fourier phase distribution produces a time series which has more extreme values with respect to a Gaussian distribution \citep{maiwald2008surrogate}. 

In the context of a turbulent signal, the Fourier phase distributions are almost uniform, and therefore, one can replace the phase values from a distribution which differs from a uniform one. Note that this procedure is not identical to phase randomization as in that case the Fourier phases are randomly shuffled without changing their distribution. Contrarily, in phase alteration experiment, we maintain the rank-wise order of the Fourier phases while sampling them from a distribution other than the original one. For our purposes, we chose von-Mises distribution to sample the Fourier phases \citep{best1979efficient}. This distribution is defined by a parameter $k$, whose value when zero indicates a uniform distribution. However, for $k>0$, the von-Mises distribution becomes increasingly different from a uniform one. Since there is no upper bound on $k$, we restricted the $k$ parameters between 0 to 9.

We apply this phase alteration technique on the $u^{\prime}$ signal at $z^{+}=$ 67 and the results are presented in Fig. \ref{fig:8}c--d. From Fig. \ref{fig:8}c, no change in $P(N_{p})$ can be seen as the $k$ parameter is varied, but the tails of $P(u^{\prime}/\sigma_{u})$ become significantly heavier than a Gaussian one (Fig. \ref{fig:8}d). Therefore, it becomes evident that by increasing $k$ more importance is given to the extreme events in the time series.  

\section{$u$, $w$ spectra and $u$-$w$ cospectra}
\label{app_B}

\begin{figure}[h]
\centering
\hspace*{-1.5in}
\includegraphics[width=1.6\textwidth]{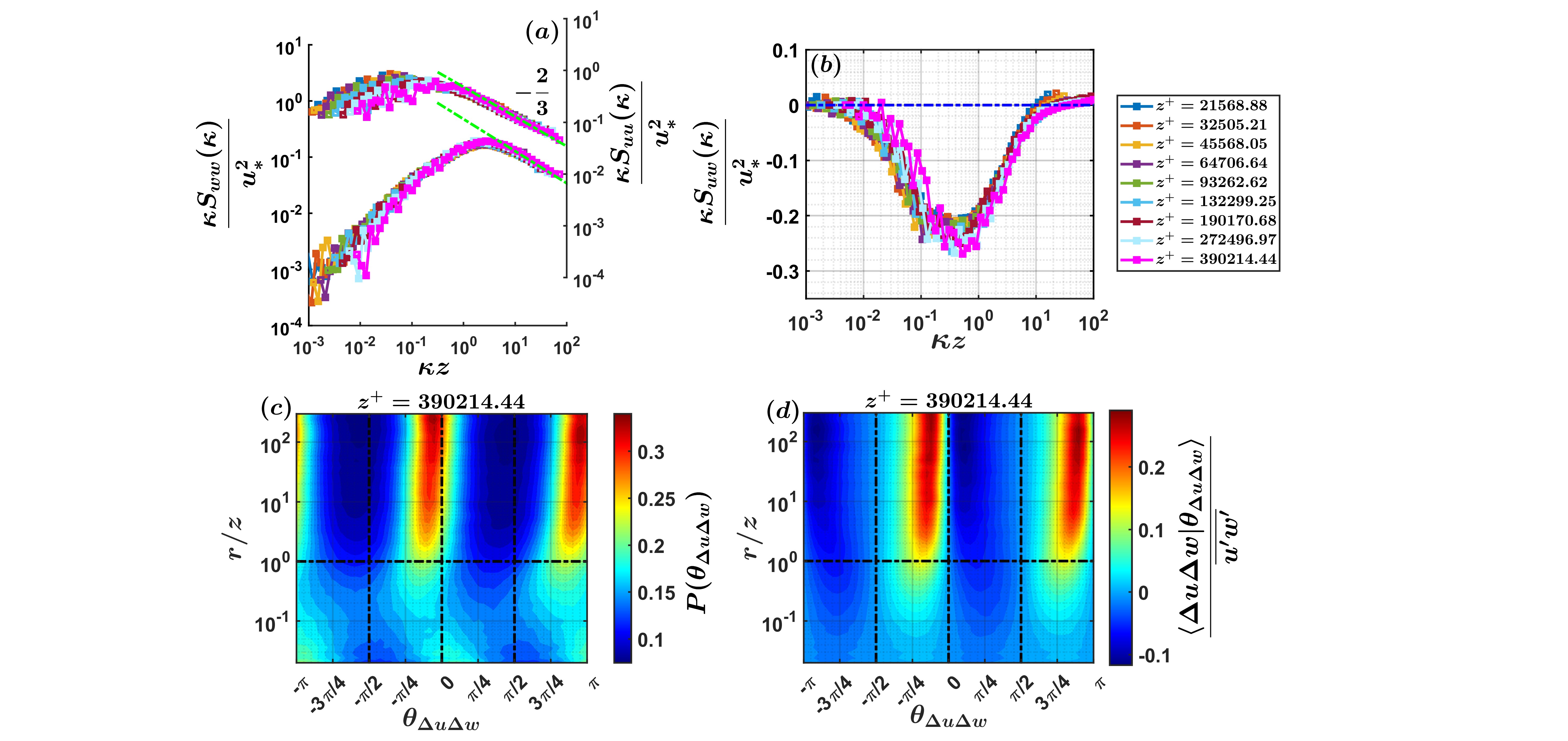}
  \caption{The averaged premultiplied (a) $u$ and $w$ spectra and (b) $u$-$w$ cospectra are shown from the near-neutral SLTEST dataset. In (a), the right-hand-side axes representing the $u$ spectra are vertically shifted for visualization purposes. The green dash-dotted lines indicate the $-2/3$ spectral slope. For both (a) and (b), spectral (cospectral) amplitudes ($\kappa S_{uu}(\kappa)$, $\kappa S_{ww}(\kappa)$, $\kappa S_{uw}(\kappa)$) are normalized by $u_{*}^2$ and the wavenumbers ($\kappa$) by height $z$. In (c) and (d), the contour plots of the probability density functions of scale-dependent phase angles ($P(\theta_{\Delta u \Delta w})$) and conditional contributions to the momentum fluxes ($\langle \Delta u \Delta w \vert \theta_{\Delta u \Delta w} \rangle$) are shown from the topmost SLTEST height. The length scales ($r$) are normalized by $z$.}
\label{fig:9}
\end{figure}

Apart from the second- and mixed-order structure functions (Fig. \ref{fig:4}d--f), we also provide the spectra of velocity fluctuations and momentum flux cospectra from the SLTEST dataset. For instance, in Fig. \ref{fig:9}a--b, the premultiplied spectra of horizontal and vertical velocity fluctuations ($\kappa S_{xx}(\kappa)$, where $x=u,w$) and the associated momentum flux cospectra ($\kappa S_{uw}(\kappa)$) are plotted against the streamwise wavenumbers ($\kappa$). These results are averaged over all the selected near-neutral runs. 

The wavenumbers ($\kappa$) are estimated by converting the frequencies to wavelengths through Taylor's hypothesis and subsequently normalized by the height above the surface ($z$). On the other hand, the spectral and cospectral amplitudes are normalized by the friction velocity ($u_{*}$). Although in the inertial subrange both $u$ and $w$ spectra display $-2/3$ slope, their behaviors are significantly different at larger scales of the flow. For instance, the $u$ spectra show a flatter region (thereby representing the $\kappa^{-1}$ scaling) while the $w$ spectral slopes are nearly equal to $+1$. Moreover, the $w$ spectral peaks reside at $\kappa z=$ 2.5. Regarding the momentum flux cospectra, they collapse nicely under the $z$ and $u_{*}$ scaling with a peak at around $\kappa z=$ 0.4.

To connect the scale-dependent momentum flux features with the coherent structures (ejections and sweeps), a polar quadrant analysis is undertaken where the phase angles and amplitudes are computed at each specific scale of the flow. For such analysis, we use the structure function analog of momentum flux (i.e., the mixed-order velocity increments $\Delta u \Delta w$) where the time-lags ($\tau$) are connected to the eddy time or length scales ($r=\tau \times \overline{u}$). Rather than the conventional joint probability density functions, polar quadrant representation is a neat way of investigating the inter-relationships between two variables \citep{chowdhuri2020persistenceflux}. To briefly explain this procedure, for each normalized scale $r/z$, one evaluates the phase angles associated with the instantaneous values of $\Delta u \Delta w$ as,
\begin{equation}
\theta_{\Delta u \Delta w}=\arctan{(\Delta w/\Delta u)}
\label{flux_vector}.
\end{equation}
Note that $\theta_{\Delta u \Delta w}$ varies between $-\pi$ to $\pi$ and these ranges are directly related to the counter-gradient ($\Delta u >0, \Delta w >0$ or $\Delta u<0, \Delta w<0$) and co-gradient motions ($\Delta u >0, \Delta w <0$ or $\Delta u < 0, \Delta w > 0$) at each scale. For instance, when $-\pi/2<\theta_{\Delta u \Delta w}<0$ or $\pi/2<\theta_{\Delta u \Delta w}<\pi$, they represent the co-gradient motions (ejections and sweeps) while the other ranges correspond to the counter-gradient ones (outward- and inward-interactions). As a consequence, the PDFs of $\theta_{\Delta u \Delta w}$ ($P(\theta_{\Delta u \Delta w})$) provide useful information about what type of motions statistically dominate the momentum transport at each scale.

Apart from $\theta_{\Delta u \Delta w}$, the momentum fluxes associated with the phase angles can be computed as,
\begin{equation}
\langle \Delta u \Delta w \vert \{\theta_{\Delta u \Delta w}(i)<\theta_{\Delta u \Delta w}<\theta_{\Delta u \Delta w}(i)+d\theta_{\Delta u \Delta w}\} \rangle=\frac{\sum \Delta u (i) \Delta w (i)} {N \times d\theta_{\Delta u \Delta w}},
\label{Flux_theta}
\end{equation}
where $i$ is the bin index, $d\theta_{\Delta u \Delta w}$ is the bin width, and $N$ is the number of samples at lags $r/z$. The division by $N$ and $d\theta_{\Delta u \Delta w}$ ensure that when integrated over $\theta_{\Delta u \Delta w}$, it would yield $\overline{\Delta u \Delta w}$ which is simply the averaged momentum flux at scale $r/z$. For our purposes, the variable at the left-hand-side of Eq. (\ref{Flux_theta}) is denoted as $\langle \Delta u \Delta w \vert \theta_{\Delta u \Delta w} \rangle$ and further normalized by the time-averaged momentum flux $\overline{u^{\prime}w^{\prime}}$. Therefore, the scale-dependent aspects of momentum flux transport can be studied more rigorously by examining this normalized quantity along with $P(\theta_{\Delta u \Delta w})$.

In Fig. \ref{fig:9}c--d, we show the contour plots of $P(\theta_{\Delta u \Delta w})$ and $\langle \Delta u \Delta w \vert \theta_{\Delta u \Delta w} \rangle / \overline{u^{\prime}w^{\prime}}$ from the topmost SLTEST height (pink lines in Fig. \ref{fig:9}a--b). We obtain identical results if any other heights were used from the SLTEST experiment. One can immediately notice, at scales $r/z > 1$, the momentum transport is mainly governed by the co-gradient motions, as the contours show their peak values at those ranges of $\theta_{\Delta u \Delta w}$. On the contrary, at inertial-subrange scales ($r/z<1$), no such clear preference towards the co-gradient motions can be noticed. Therefore, the bulk of the momentum flux are transported through the ejection and sweep motions at scales commensurate with the energy-production scales. 

\section{PDFs of velocity fluctuations and momentum flux}
\label{app_C}

\begin{figure}[h]
\centering
\hspace*{-1.7in}
\includegraphics[width=1.6\textwidth]{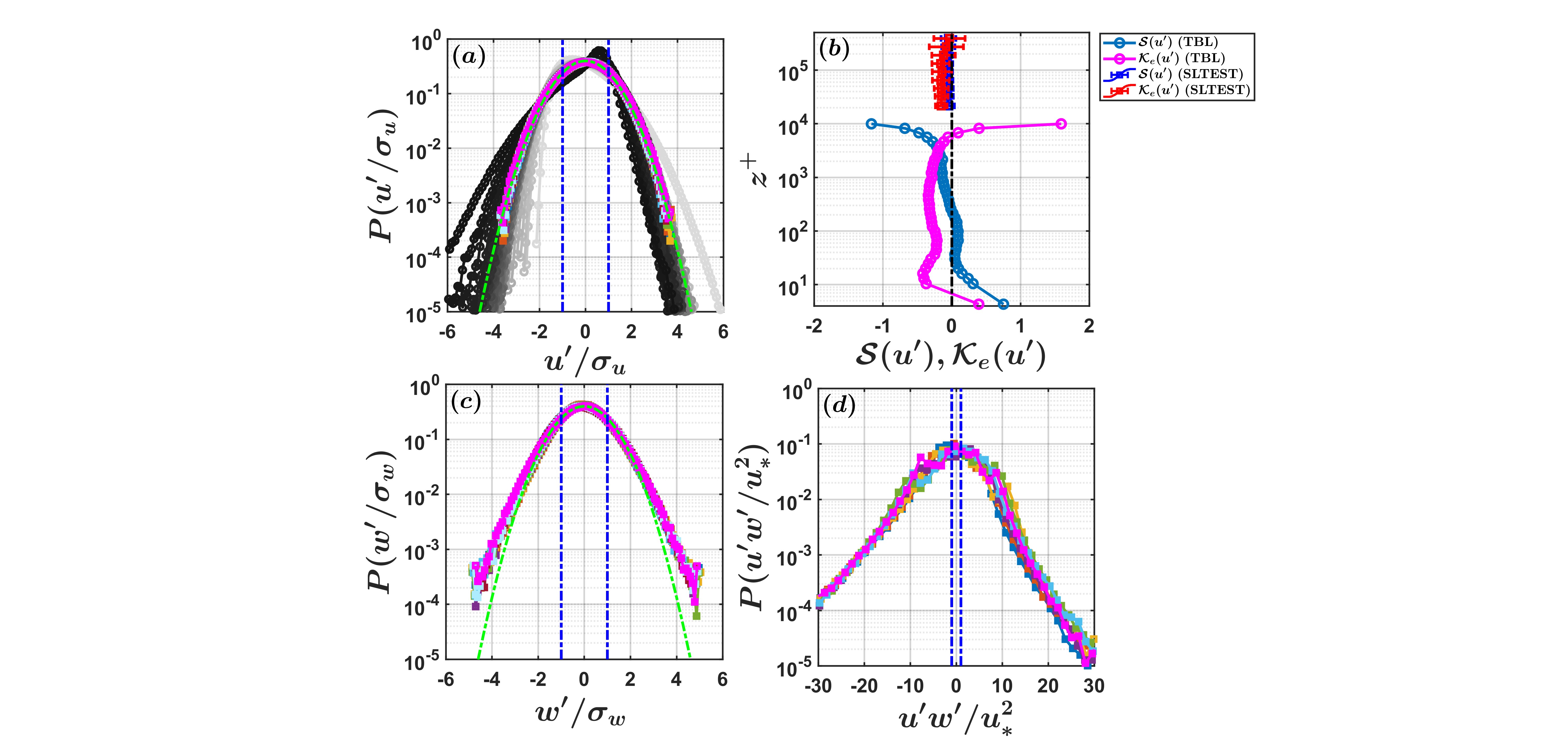}
  \caption{(a) The PDFs of $u^{\prime}/\sigma_{u}$ are shown from the TBL and SLTEST experiments. (b) The skewness ($\mathcal{S}$) and excess kurtosis ($\mathcal{K}_{e}$) of the $u^{\prime}$ signals are plotted against $z^{+}$ (see the legend). The error-bars denote the spread around the mean for the SLTEST dataset. The PDFs of (c) $w^{\prime}/\sigma_{w}$, and (d) $u^{\prime}w^{\prime}/u_{*}^2$ are displayed from the SLTEST experiments. The green colored lines in (a) and (c) indicate the Gaussian distribution. The color codes are similar to the legend in Fig. \ref{fig:4}. Two blue dash-dotted lines in (a), (c), and (d) highlight the values $\pm$ 1 in order to emphasize the importance of the large amplitude events in respective signals.}
\label{fig:10}
\end{figure}

The bursts in a signal are typically characterized through their PDFs. In Fig. \ref{fig:10}, we describe the PDFs of streamwise and vertical velocity fluctuations ($u^{\prime}$ and $w^{\prime}$) and instantaneous momentum flux ($u^{\prime}w^{\prime}$) signals. The quantities $u^{\prime}$ and $w^{\prime}$ are normalized with their respective standard deviations ($\sigma_{u}$ and $\sigma_{w}$). On the other hand, $u^{\prime}w^{\prime}$ signals are normalized with $u_*^2$. By comparing the PDFs of $u^{\prime}$, a difference is noted between the two experiments (Fig. \ref{fig:10}a). For instance, although the PDFs of $u^{\prime}$ from the SLTEST experiment are strictly Gaussian at all the levels (a nice collapse is evident), a deviation from Gaussianity is observed for the TBL experiment. This is highlighted through the vertical profiles of skewness ($\mathcal{S}$) and excess kurtosis ($\mathcal{K}_e$) in Fig. \ref{fig:10}b. Note that $\mathcal{K}_e$ is obtained after subtracting 3 of a Gaussian distribution. 

In addition to $u^{\prime}$, the normalized PDFs of $w^{\prime}$ and $u^{\prime}w^{\prime}$ collapse nicely for the SLTEST experiment (Figs. \ref{fig:10}c--d). The PDFs of $w^{\prime}$ display a heavier tail towards the positive values, while the PDFs of $u^{\prime}w^{\prime}$ remain skewed towards the negative side. From Fig. \ref{fig:10}d, one can notice that $P(u^{\prime}w^{\prime}/u_*^2)$ show heavy tails beyond $\pm$ 1, thereby indicating the presence of extreme flux events (significantly larger than the mean flux values) at all the nine SLTEST levels.

\end{appendices}
\bibliographystyle{apalike}  
\bibliography{references}
\clearpage

\section*{Supplementary material}
\renewcommand{\thefigure}{S\arabic{figure}}
\setcounter{figure}{0}

\begin{figure*}[h]
\centering
\hspace*{-1.8in}
\includegraphics[width=1.5\textwidth]{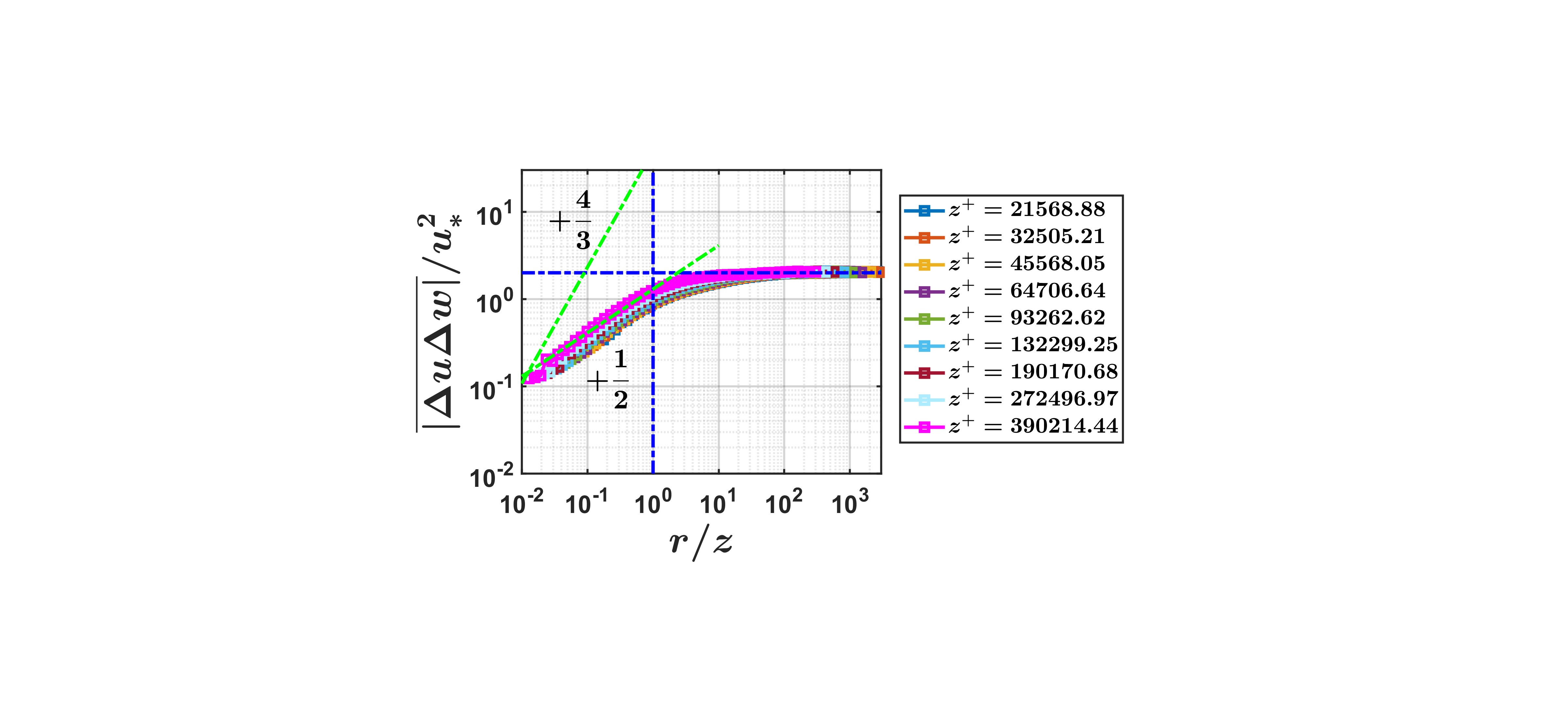}
\caption{The averages of absolute mixed-order velocity increments ($\overline{|\Delta u \Delta w|}/u_{*}^2$, normalized with $u_{*}$) are plotted against $r/z$ for the near-neutral SLTEST dataset. Due to taking absolute values, in the inertial-subrange scales, a $+1/2$ scaling is observed as opposed to $+4/3$.}
\end{figure*}

\begin{figure*}[h]
\centering
\hspace*{-1.8in}
\includegraphics[width=1.5\textwidth]{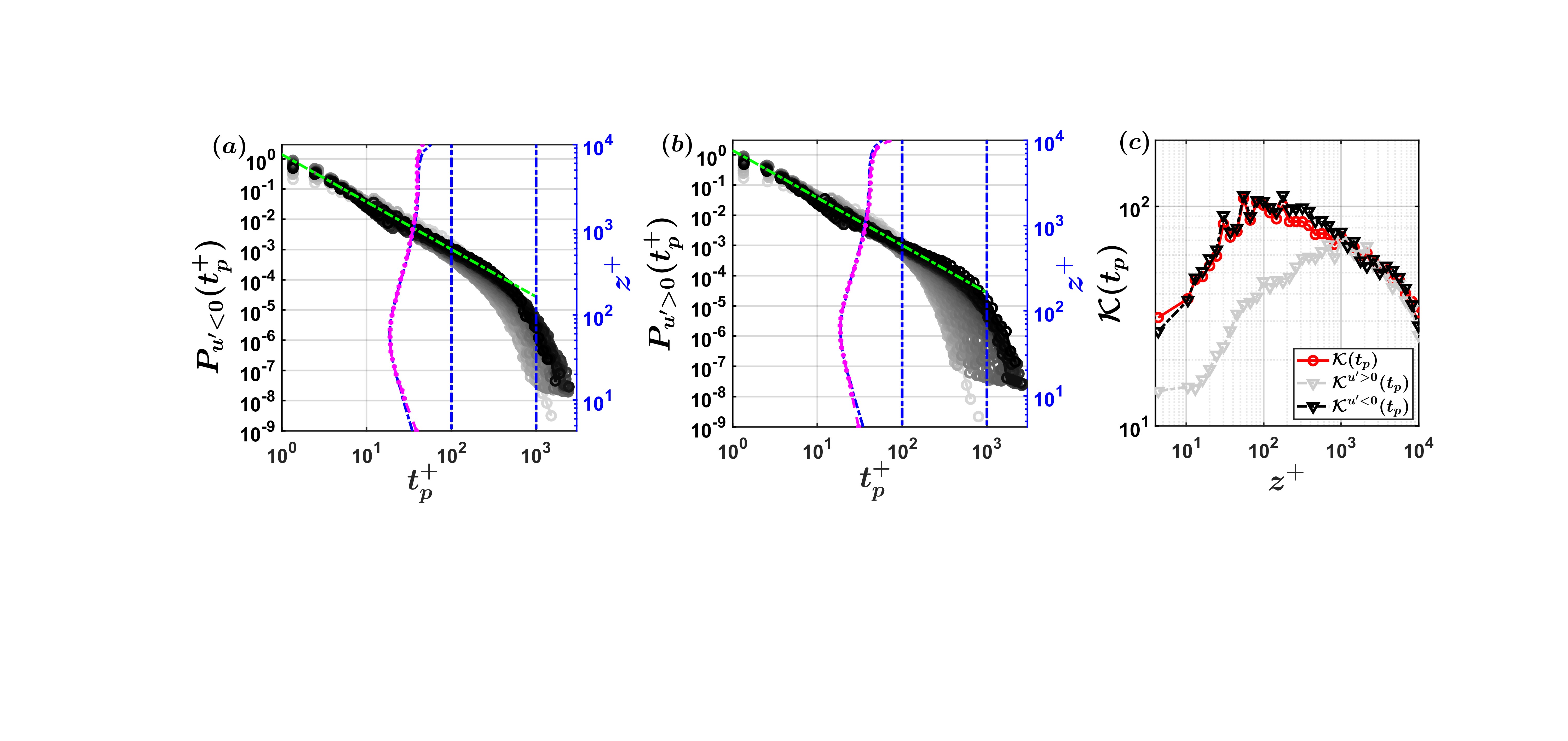}
\caption{The PDFs of normalized event time scales (${t_{p}}^{+}$) are shown from the TBL dataset, corresponding to (a) $u^{\prime}<0$ and (b) $u^{\prime}>0$ signals at various $z^{+}$ locations. The two dash-dotted blue vertical lines in (a) and (b) represent the inner- and outer-spectral peak positions evaluated from the $u$ spectra. For both (a) and (b), the right-hand-side of the $y$ axis represents the vertical profiles of mean event time scales ($\overline{{t_{p}}^{+}}$). The pink lines with markers in (a) and (b) indicate $\overline{{t_{p}}^{+}}$ by considering both the positive and negative events in $u^{\prime}$ signal. On the other hand, the blue lines with markers show $\overline{{t_{p}}^{+}}$ computed from the (a) $u^{\prime}<0$ and (b) $u^{\prime}>0$ signals, respectively. In (c), we show the variations of the kurtosis of $t_{p}$ ($\mathcal{K}(t_{p})$) with $z^{+}$, plotted separately for $u^{\prime}$, $u^{\prime}>0$, and $u^{\prime}<0$  signals (see the legend).}
\end{figure*}

\begin{figure*}[h]
\centering
\hspace*{-1.8in}
\includegraphics[width=1.6\textwidth]{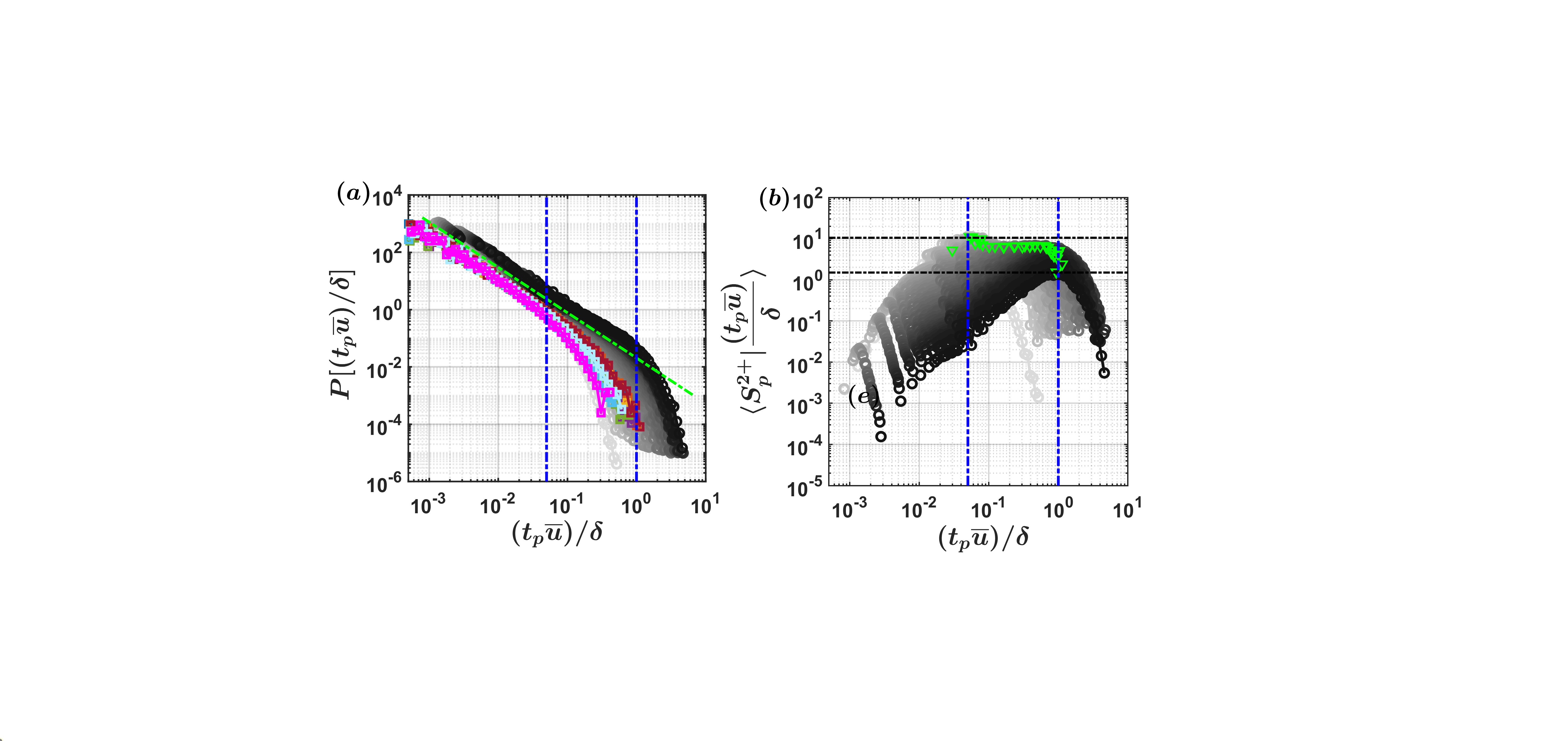}
\caption{(a) The PDFs of event length scales ($t_{p}\overline{u}$), normalized with the boundary-layer depth ($\delta$), are shown from both the TBL and SLTEST datasets. For the SLTEST dataset, the direct estimation of $\delta$ was not available, and therefore, the integral scale of $u^{\prime}$ signal at $z=25.69$ m is used instead. The green dash-dotted line denotes the $-1.6$ power-law. (b) The normalized event contributions to the velocity variances are plotted against $t_{p}\overline{u}/\delta$ for various $z^{+}$ locations from the TBL dataset. The green markers denote the peak positions of the event contribution curves. The two horizontal black dash-dotted lines represent the maximum and minimum contour levels in Fig. 5c of the paper. For both (a) and (b), the vertical blue dash-dotted lines indicate the $\delta$-scaled inner- and outer-spectral peak positions.} 
\end{figure*}

\begin{figure*}[h]
\centering
\hspace*{-2.2in}
\includegraphics[width=1.6\textwidth]{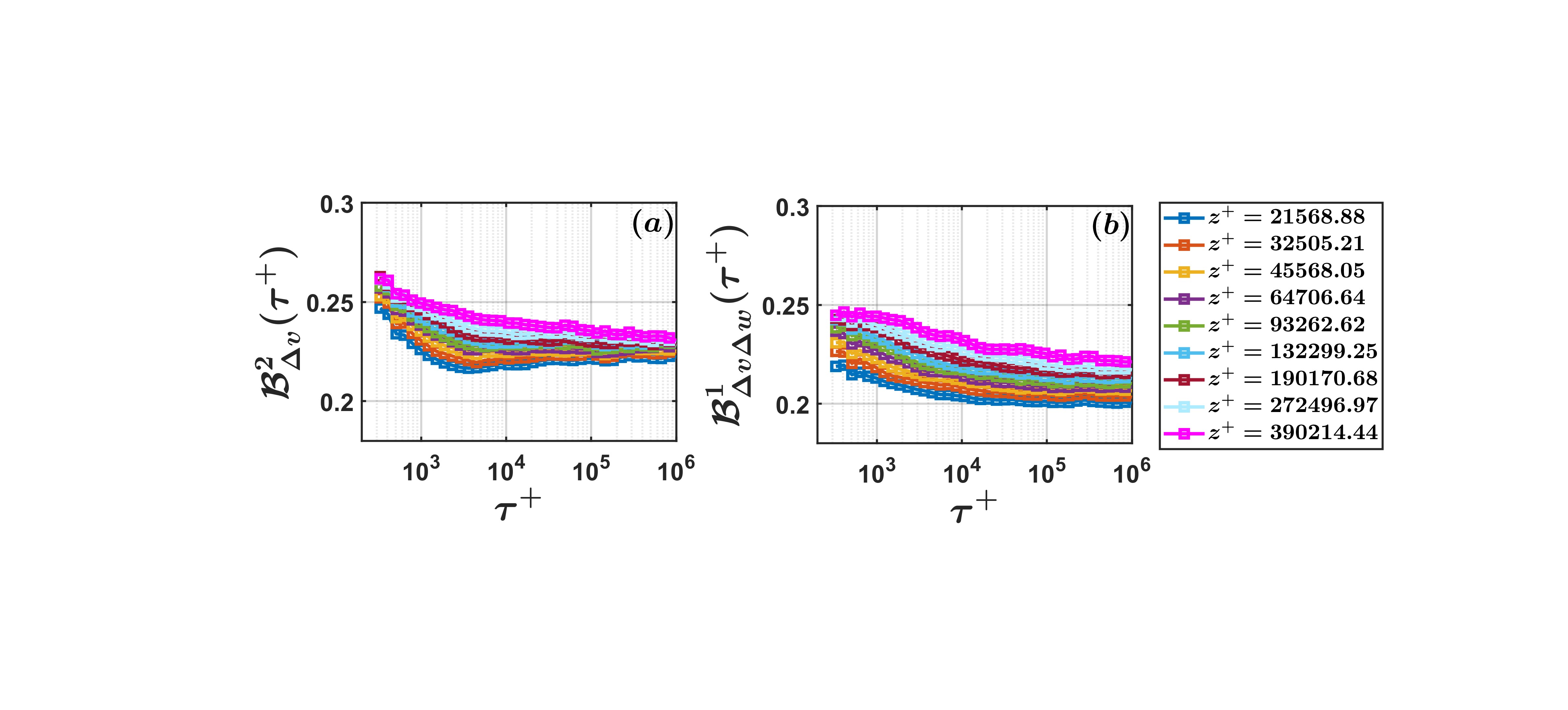}
\caption{A scale-wise evolution of burstiness indices are plotted for (a) ${\Delta v}^2$ ($B^{2}_{\Delta v}(\tau^{+})$) and (b) $\Delta v \Delta w$ ($B^{1}_{\Delta v \Delta w}(\tau^{+})$) signals against the normalized time lags $\tau^{+}$. Different SLTEST heights are shown in the legend.} 
\end{figure*}

\begin{figure*}[h]
\centering
\hspace*{-1.6in}
\includegraphics[width=1.6\textwidth]{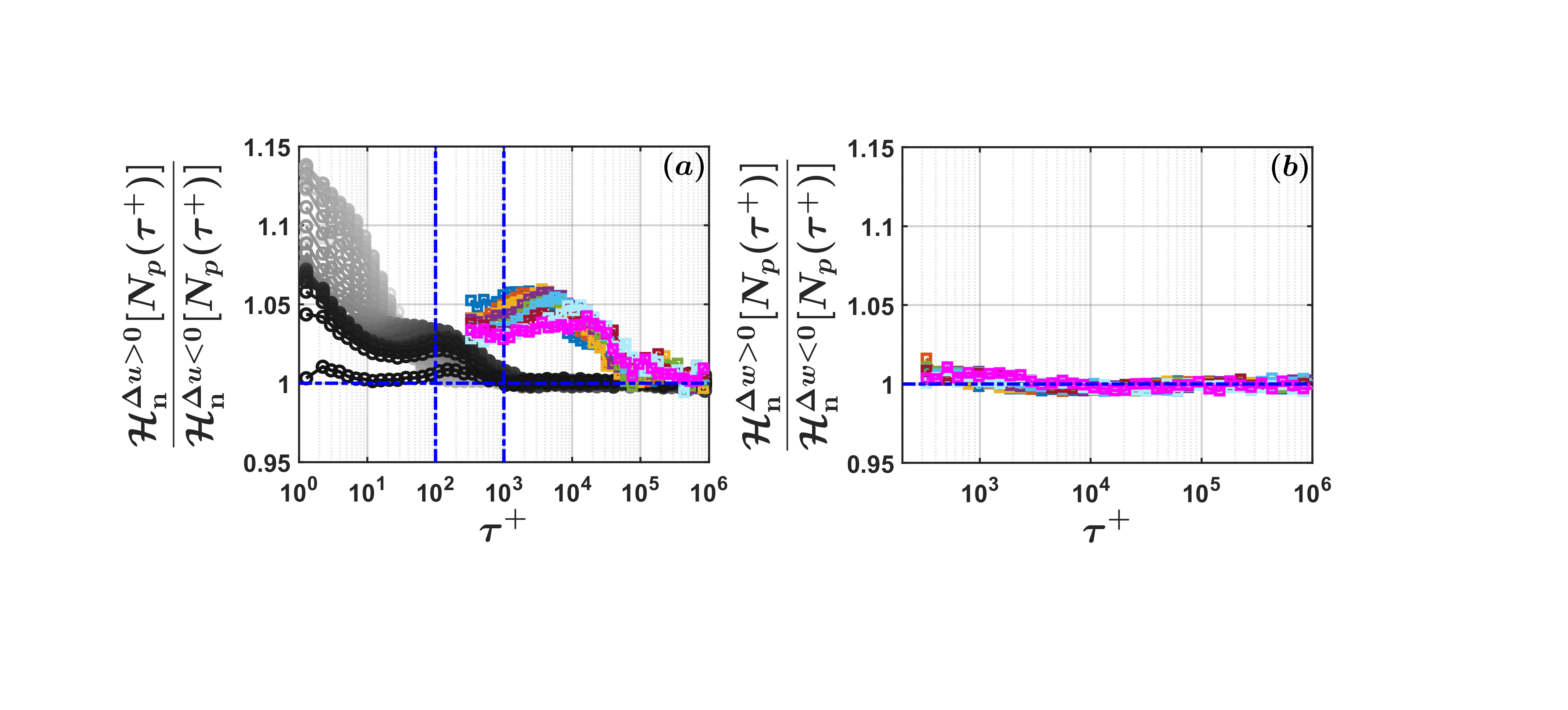}
\caption{The entropy ratios between the positive and negative velocity increments are plotted for the (a) $\Delta u$ and (b) $\Delta w$ signals, corresponding to the TBL and SLTEST datasets. The color schemes are same as the previous figures.} 
\end{figure*}

\end{document}